\documentclass[aps, floatfix, twocolumn, a4paper, showpacs, nofootinbib,
superscriptaddress, 10pt]{revtex4}
\usepackage{graphicx,float}
\usepackage[all]{xy}
\usepackage{amsmath,upgreek}
\usepackage{amssymb}
\usepackage{color}
\usepackage{epsfig}		
\usepackage{graphicx,epstopdf}
\usepackage{subfigure}
\usepackage{pdfpages}
\usepackage[colorlinks,hyperindex]{hyperref}

\definecolor{green1}{RGB}{0,128,0} 
\hypersetup{backref=true,pagebackref=true}
\hypersetup{%
  colorlinks = true,
  linkcolor  = blue,
  citecolor = cyan,
}
\usepackage{bookmark,textgreek}
\usepackage{hyperref,color,xcolor}
\hypersetup{hidelinks,hyperindex=true,colorlinks=true,breaklinks=true,urlcolor= blue}
\hypersetup{%
  colorlinks = true,
  linkcolor  = blue
}
\usepackage{graphicx,float,tikz}
\usepackage[all]{xy}
\newcommand\ringring[1]{%
  {% make an Ord atom
   \mathop{\kern0pt #1}\limits^{% set a box over the variable
     \vbox to-1.85ex{
       \kern-2ex % lower the ring accents
       \hbox to 0pt{\hss\normalfont\kern.1em \r{}\kern-.45em \r{}\hss}%
       \vss % fill
     }% end of \vbox
   }% end of the superscript
  }% end of \mathop
}
\usepackage[normalem]{ulem} %pra usar o \sout{...}

\newcommand\orcidroldao{{\href{https://orcid.org/0000-0003-3978-532X}{\orcidicon}}}
\newcommand{\orcidicon}{%
	\begin{tikzpicture}
	\draw[lime, fill=lime] (0,0)
		circle [radius=0.16]
		node[white] {{\fontfamily{qag}\selectfont \tiny ID}};
	\draw[white, fill=white] (-0.0625,0.095)
		circle [radius=0.007];
	\end{tikzpicture}	\hspace{-2mm}
}

\newcommand{\bes}{\begin{subequations}}
\newcommand{\ees}{\end{subequations}}
\def\beq{\begin{eqnarray}}
 \newcommand{\cclt}{\textcolor{black}}
 \newcommand{\blt}{\textcolor{black}}
  
\def\eeq{\end{eqnarray}}
\def\be{\begin{equation}}
\def\ee{\end{equation}}

 \newcommand{\aalt}{\textcolor{black}}
 
\begin{document}

\title{Configurational entropy and spectroscopy of \blt{even-spin glueball} resonances in dynamical AdS/QCD}

\author{D. Marinho Rodrigues}
\email{diegomhrod@gmail.com}
\affiliation{Federal University of ABC, Center of Mathematics, Santo Andr\'e, 09580-210, Brazil}
\affiliation{Federal University of ABC, Center of Physics, Santo Andr\'e, 09580-210, Brazil}
\author{R. da Rocha\orcidroldao\!\!}
\email{roldao.rocha@ufabc.edu.br}
\affiliation{Federal University of ABC, Center of Mathematics, Santo Andr\'e, 09580-210, Brazil}
\affiliation{}
\begin{abstract}
{\begin{center}{\bf Abstract}\end{center}}
\blt{Even-spin glueball} resonances in AdS/QCD are here studied using the configurational entropy (CE). The concept of CE Regge trajectories, associating the CE of the \blt{even-spin glueball} resonances with both their spin $J^{PC}$ and to their mass spectra, is used to derive the mass spectra of higher $J^{PC}$ resonances with $J$ even. For it, the linear, the exponential modified, and the anomalous quadratic dilatonic models, each one with linear and logarithmic anomalous corrections, are employed. 
Several methods are implemented, hybridizing AdS/QCD and established data of lattice QCD.  
\end{abstract}
\maketitle

\section{Introduction}

The configurational entropy (CE) displays the compression rate of all information flowing through a system, in the limit where no losses are regarded. Information is compressed into momentum space modes, that comprise a given physical system, as firstly proposed by Shannon \cite{Shannon:1948zz}. In this sense, the CE constitutes a bound on the transmission rate of information sources. \blt{In Ref. \cite{Shannon:1948zz}, Shannon promoted information, from a previous concept that had not been established yet, to a very precise theory, consisting of the theory underlying any present-day digital setup. Shannon's seminal work was the foundation of developing algorithms with data-compression codes that are error-correcting, likewise. To estimate the information that underlies the system, a probability distribution is employed using the momentum space. In this scenario, distinct wavelengths enforce the creation of correlations among all the system constituents \cite{Gleiser:2011di,Fernandes-Silva:2019fez,Gleiser:2018kbq,roldao,Bernardini:2016hvx,Gleiser:2012tu}.}
The CE can probe and also measure the information that composes the correlations among these modes, thus converting into a coded form the complexity of shape in the studied system \cite{Gleiser:2011di,Gleiser:2012tu}. \blt{Both informational and dynamical aspects of the studied system are interplayed by the CE \cite{Gleiser:2012tu}.
The CE was also employed to discern the set of dynamical systems among chaotic, random, and regular, also evaluating the chaotic progression in several QCD setups. This method brings additional tools to make out physical phenomena in QCD \cite{Ma:2018wtw}.} Localized solutions of PDEs are examples of non-trivial spatial complexity, as well as the correlation of thermal fluctuations in systems that go through phase transitions. For computing the CE, a spatially localized integrable scalar field is necessary. The energy density, taken as the time component of the energy-momentum tensor, is a prime scalar field candidate \cite{Sowinski:2015cfa}. However, other localized scalar fields, like the nuclear cross-section and scattering amplitudes have been used 
 in different QCD contexts \cite{Karapetyan:2016fai,Karapetyan:2017edu,Karapetyan:2018oye,Karapetyan:2018yhm}. 
%Further aspects of the CE were presented in Ref. \cite{Gleiser:2018kbq}. 
The CE has been demonstrated to consist of a very useful apparatus to study and predict the prevalence, abundance, and dominance of physical states and their resonances. The recently obtained results, in the literature, mainly in QCD and AdS/QCD setups, match and corroborate to phenomenological data in PDG \cite{pdg1}, as well as
provide relevant predictions for future experiments, mainly in LHC. 

The CE has been playing important role in scrutinizing AdS/QCD (both hardwall and softwall) models, studying relevant properties in QCD, and its phenomenology. Several new methods and procedures, also involving the existence of CE Regge trajectories, have been employed to investigate diverse families of light-flavor mesonic states and excitations, in chiral-gluon condensates and dilaton-graviton backgrounds \cite{daRocha:2021ntm,daRocha:2021imz,Bernardini:2016hvx,Bernardini:2018uuy,Ferreira:2019inu}, also including tachyonic ones \cite{Barbosa-Cendejas:2018mng}. Besides, 
tensor mesons \cite{Ferreira:2019nkz} and 
scalar glueballs \cite{Bernardini:2016qit} were explored under CE tools.  Quarkonia states, both at zero temperature \cite{Braga:2017fsb} and at finite temperature, \cite{Braga:2018fyc}, were investigated, with relevant physical features regarding phenomenology and the configurational stability of bottomonium and charmonium resonances. Besides, quarkonia and plasmas with finite density were studied, from the CE point of view, in Ref. \cite{Braga:2020myi}. Refs. \cite{Karapetyan:epjp,Karapetyan:2021epjp,Karapetyan:2016fai,Karapetyan:2017edu,Karapetyan:2018oye,Karapetyan:2018yhm,Karapetyan:2019ran,daSilva:2017jay} studied gluons and quarks in the color glass condensate regime of QCD, exhibiting new features and applications of CE, and providing a theoretical explanation for diverse parameters, previously used in the literature to best fit experimental data. More on the use of CE in QCD was studied in Ref. \cite{Ma:2018wtw}. Standard and exotic baryonic excitations were scrutinized in Ref. \cite{Colangelo:2018mrt}, also in a finite temperature setup \cite{Ferreira:2020iry}. 
The configurational stability of boson stars and black holes were also explored, using the CE, in Refs. \cite{Casadio:2016aum,Gleiser:2015rwa,Braga:2016wzx,Braga:2019jqg,Lee:2018zmp,Lee:2017ero}.
%,Fernandes-Silva:2019fez}, 
Also, phase transitions \cite{Gleiser:2018kbq}, topological defects \cite{Correa:2016pgr,Cruz:2019kwh,Lee:2019tod,Bazeia:2018uyg,Bazeia:2021stz} and aspects of particle physics \cite{Alves:2014ksa,Alves:2017ljt} were discussed with the use of the CE. 
The quantum version of the classical Shannon entropy is detailedly exposed in Ref. \cite{Witten:2018zva}.
  
The AdS/QCD setup has been proved to be an important tool to investigate non-perturbative aspects of QCD, as, for example, confinement \cite{EKSS2005,Karch:2006pv,Brodsky:2014yha,Maldacena:1997re,Witten:1998qj,gub}. The hard and the softwall models represent thriving approaches of AdS/QCD that match data in PDG \cite{Polchinski:2001tt,BoschiFilho:2002ta,BoschiFilho:2002vd,Huang:2007fv}. In the gauge/gravity duality dictionary, fields in the AdS$_5$ bulk represent dual quantities to operators of QCD on the boundary. %\cite{Natsuume:2014sfa}. 

Several phenomenological approaches to scrutinize non-perturbative features of QCD consist in probing particle states, and their resonances. QCD predicts gluonic bound states, comprising glueballs, to exist. Therefore the quest for the glueball content of hadrons plays an important role. One of these approaches is based on the construction of phenomenological Lagrangians capable of taking into account gluonic processes in hadronic interactions \cite{Gutsche:2016wix}.~Lattice QCD computations derived the glueball spectrum in many setups \cite{Meyer:2004jc,Sergeenko:2011kf,Donoghue:1980hw}. 
When one employs solutions of the equations of motion for massive gluons, glueball masses can be derived, in full compliance with QCD lattice phenomenology. Then, the complex pomeron trajectory can be obtained, whose real projection regards the soft pomeron, with support of HERA data \cite{Sergeenko:2011kf}. 
Among prominent topics in the physics of hadrons, the relationship between the pomeron and \blt{even-spin} glueballs plays a prominent role \cite{Ewerz:2013kda}. 
Our main aim here is to use techniques of CE, already well succeeded in QCD applications, to study \blt{even-spin glueball} resonances, also in the context of AdS/QCD. Configurational-entropic Regge trajectories are 
the main tool to derive the mass spectra of higher \blt{even-spin glueball} resonances, in different dilaton models. Both AdS/QCD and lattice QCD data are going to be used, for it.  
This paper is organized as follows: Sect. \ref{sec2} briefly reviews the holographic setup used in this work, namely, the background geometry and glueball spectra. Sect. \ref{ce1} discusses the results concerning the CE applied to \blt{even-spin glueball} resonances, 
including a comprehensive analysis of the CE Regge trajectories and the derivation of the mass spectra of \blt{even-spin glueball} family resonances, in six dilaton models. The conclusions and more discussion are comprised of Sect. \ref{sec3}. 

\section{Holographic Setup}
\label{sec2}

Here the holographic model in Refs. \cite{Li:2013oda, FolcoCapossoli:2016ejd}, based on the dynamical softwall model, will be used. Next, the results for the quadratic dilaton profile are reviewed. Then, this holographic model is applied to the linear and exponential dilaton profiles.
The dynamical softwall model is characterized by an Einstein-dilaton action, in the Einstein frame, given by
\begin{equation} \label{action}
\!\!\!\!\!\!S \!=\! \dfrac{1}{16\pi G_5}\int d^{5}x \sqrt{-g}\left(R \!-\! \dfrac{4}{3}g^{\mu\nu}\partial_{\mu}\phi\partial_{\nu}\phi \!+\! V(\phi)\right),
\end{equation}
where $ G_5 $ is the 5-dimensional Newton's constant, $g=\mathrm{det}(g_{\mu\nu})$, $ R $ denotes the Ricci scalar, $\phi$ is the dilaton field and $V(\phi)$ stands for the dilaton potential. The equations of motion, derived from the action \eqref{action}, are given by
\begin{eqnarray}
\!\!\!\!\!\!\!\!\!\!\!E_{\mu\nu} -\dfrac{4}{3}\left(\partial_{\mu}\phi\partial_{\nu}\phi - \dfrac{1}{2}g_{\mu\nu}(\partial \phi)^2\right) - \dfrac{1}{2}g_{\mu\nu}V(\phi) &=& 0, \label{EinsteinEqn} \\
g^{\mu\nu}\partial_{\mu}\partial_{\nu}\phi + \dfrac{3}{8}\dfrac{\partial V(\phi)}{\partial\phi}&=& 0, \label{DilatonEqn}
\end{eqnarray}
with the Einstein tensor $ E_{\mu\nu} $ defined as
\begin{equation}
E_{\mu\nu} = R_{\mu\nu} - \dfrac{1}{2}g_{\mu\nu}R.
\end{equation}
To solve the equations of motion, the metric ansatz to be considered here is given by\footnote{In this work the AdS radius $L$ is set to unity.} \cite{Ballon-Bayona:2017sxa}
\begin{eqnarray}
ds^2 &=& \dfrac{1}{\zeta(z)^2}\left(dz^2 - dt^2 + d\vec{x}^2\right), \label{Metriczeta} \\
\phi &=& \phi(z).
\end{eqnarray}
Plugging this ansatz %ou these ansatze
in Eqs. (\ref{EinsteinEqn}, \ref{DilatonEqn}) yields
\begin{eqnarray}
\frac{\zeta''(z)}{\zeta(z)} &=& \frac{4}{9}\phi'(z)^2,\label{breqn3} \\
V(\phi) &=& 12\,\zeta'(z)^2 - \frac{4}{3}\zeta (z)^2\phi'(z)^2.\label{Veqn3}
\end{eqnarray}

The advantage of this dynamical AdS/QCD model is that the background is a consistent solution of Einstein's equations and, within it, one can implement confinement, characterized by an area-law for the Wilson loop, and linear Regge trajectories (see for instance \cite{dePaula:2008fp,Li:2011hp}).

\subsection{Quadratic Dilaton Profile}

Considering the quadratic dilaton profile $ \phi(z) = \pm k\,z^2$, the solution for $\zeta(z)$, and the dilaton potential, read
\begin{eqnarray}
\!\!\!\!\!\!\!\zeta(z) &=& z\,\,_0F_{1}\left(\dfrac{5}{4};\dfrac{k^2\,z^4}{4}\right) \label{zetasol},\\
\!\!\!\!\!\!\!V(\phi) &=& 12 \,_0F_1\left(\frac{1}{4};\frac{\phi^2}{9}\right)^2-\frac{16}{3}\,\phi^2\,
_0F_1\left(\frac{5}{4};\frac{\phi^2}{9}\right)^2,\label{dilpotsol}
\end{eqnarray}
where $_0F_{1}\left(a;z\right) $ denotes the confluent hypergeometric function. The integration constants were chosen so that $ \zeta(z)\to z $ in the UV regime ($ z\to0 $). Besides, the solution for $ \zeta(z) $ can be still put into the Bessel form, given by
\begin{equation}
\zeta(z) = \left( \dfrac{3}{k}\right)^{1/4}\Upgamma\left(\dfrac{5}{4} \right)\,\sqrt{z}\,\,I_{\frac{1}{4}}\left(\dfrac{2}{3}k\,z^2 \right),
\end{equation}
where $ I_{\nu}(z) $ is the modified Bessel function of first kind.

The UV expansions for $ \zeta(z) $ and $ V(\phi) $ read \begin{eqnarray}
\zeta(z) = z + \dfrac{4}{45}\,k^2\,z^5+\mathcal{O}(z^6),\\
V(\phi) = 12 + \dfrac{16}{3}\,\phi^2+\mathcal{O}(\phi^4).
\end{eqnarray}
Therefore, one can see that, near the UV, the geometry is asymptotically AdS.

On the other hand, the IR expansions, i.e., $z\to\infty$, go like
\begin{eqnarray}
\zeta(z) \simeq e^{\frac{2}{3}\,k\,z^2} \\
V(\phi) \simeq e^{\frac{4}{3}\,\phi} .
\end{eqnarray}
Here, it is worth pointing out that, although the full solutions for $\zeta(z)$ and $V(\phi)$, respectively in Eqs. \eqref{zetasol} and \eqref{dilpotsol}, are independent of the sign of the dilaton profile, to get the proper IR behavior one should stick to the positive sign for the dilaton field. Otherwise one would violate the confinement criteria, established in Refs. \cite{Gursoy:2007cb,Gursoy:2007er}.

\subsection{Linear Dilaton Profile}

Now, considering the linear dilaton profile, $\phi(z) = \sqrt{k}\,z$, the solutions for $\zeta(z)$ and the dilaton potential respectively read
\begin{eqnarray}
\zeta(z) &=& \frac{3\,\sinh \left(\frac{2\,\sqrt{k}\,z}{3}\right)}{2\,\sqrt{k}},\label{zetalinear} \\
V(\phi) &=& \frac{3}{2}\,\left[5 + 3\,\cosh \left(\frac{4\,\phi}{3}\right)\right].\label{potlinear}
\end{eqnarray}
Similar background solutions were found in \cite{Li:2011hp} by using the warp factor as an input to obtain the corresponding dilaton profile. 
The UV expansions, for $ \zeta(z) $ and $ V(\phi)$, in this case take the form
\begin{eqnarray}
\zeta(z) &=& z + \dfrac{2}{27}\,k\,z^3+\mathcal{O}(z^5),\\
V(\phi) &=& 12 + 4\,\phi^2+\mathcal{O}(\phi^4).
\end{eqnarray}
Again, the UV regime yields the geometry to be asymptotically AdS. 
Besides, the IR expansions yield 
\begin{eqnarray}
\zeta(z) &\simeq& e^{\frac{2}{3}\,\sqrt{k}\,z} \\
V(\phi) &\simeq& e^{\frac{4}{3}\,\phi},
\end{eqnarray}
which looks very similar to the IR expansion for the quadratic dilaton case.

\subsection{Exponential Dilaton Profile}

Here, the background described by the action \eqref{action} is investigated, 
with a modified exponential dilaton profile, given by \cite{Fang:2019lmd}
\begin{equation}\label{expdilaton}
\phi(z) = k\,z^2(1-e^{-k\,z^2}),
\end{equation}
whose asymptotic behavior interpolates between the quadratic dilaton profile, $ \phi(z) = k\,z^2 $, in the IR regime, and a quartic dilaton profile, $ \phi(z) = k\,z^4 $, in the UV limit. This profile was used in Ref. \cite{Fang:2019lmd}, in the context of the meson spectra and chiral phase transitions, providing quite good results. Here we want to study the effect of this profile from the point of view of the even-spin glueball spectra, associated with the pomeron, using the CE.

\begin{figure}[h]
	\centering
	\includegraphics[scale = 0.26]{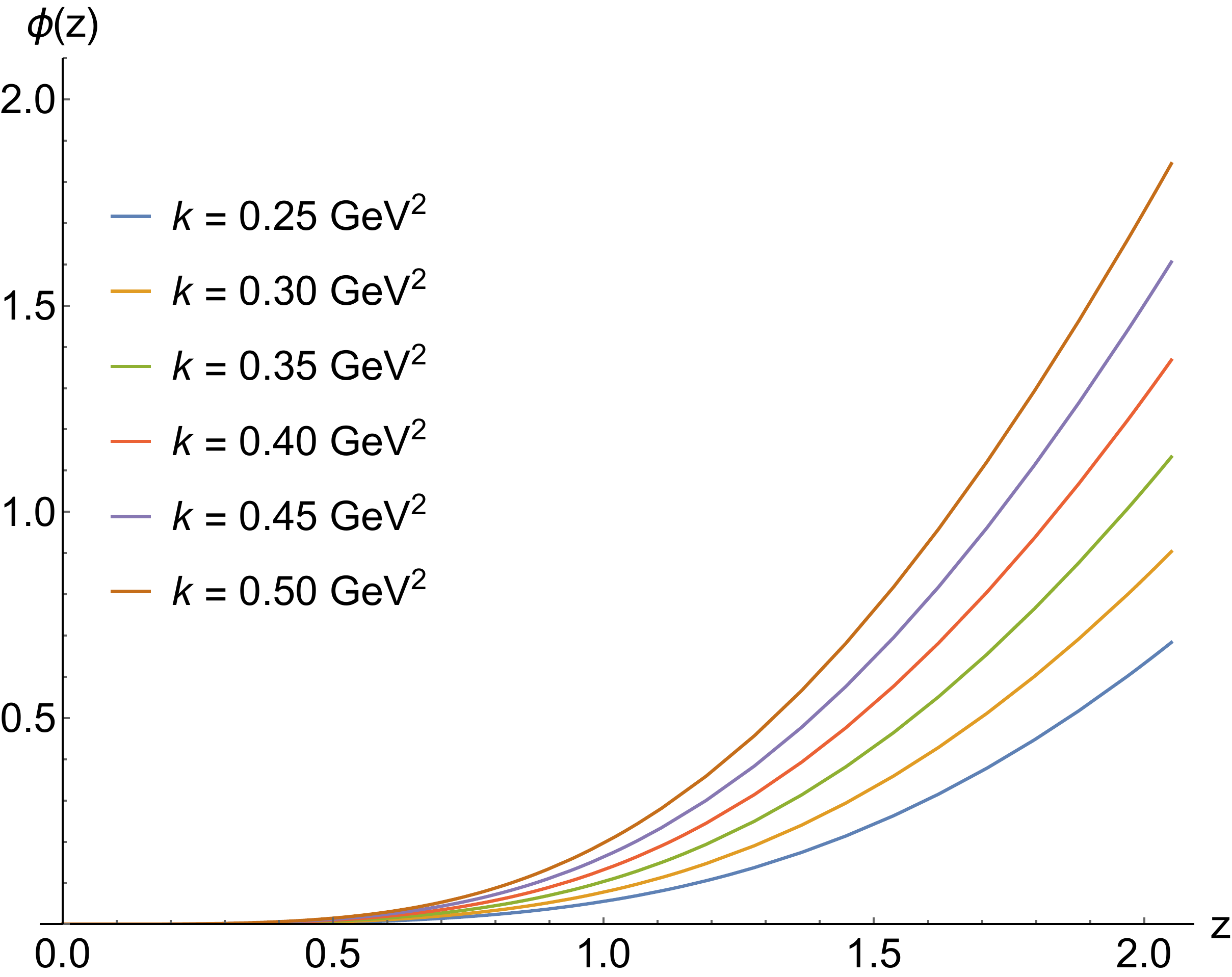}
	\caption{Exponential dilaton $\phi(z)$ \eqref{expdilaton} profile, for several values of $k$ in GeV$^2$.}
	\label{fig:dilaton}
\end{figure}

\begin{figure}[h]
	\centering
	\includegraphics[scale = 0.26]{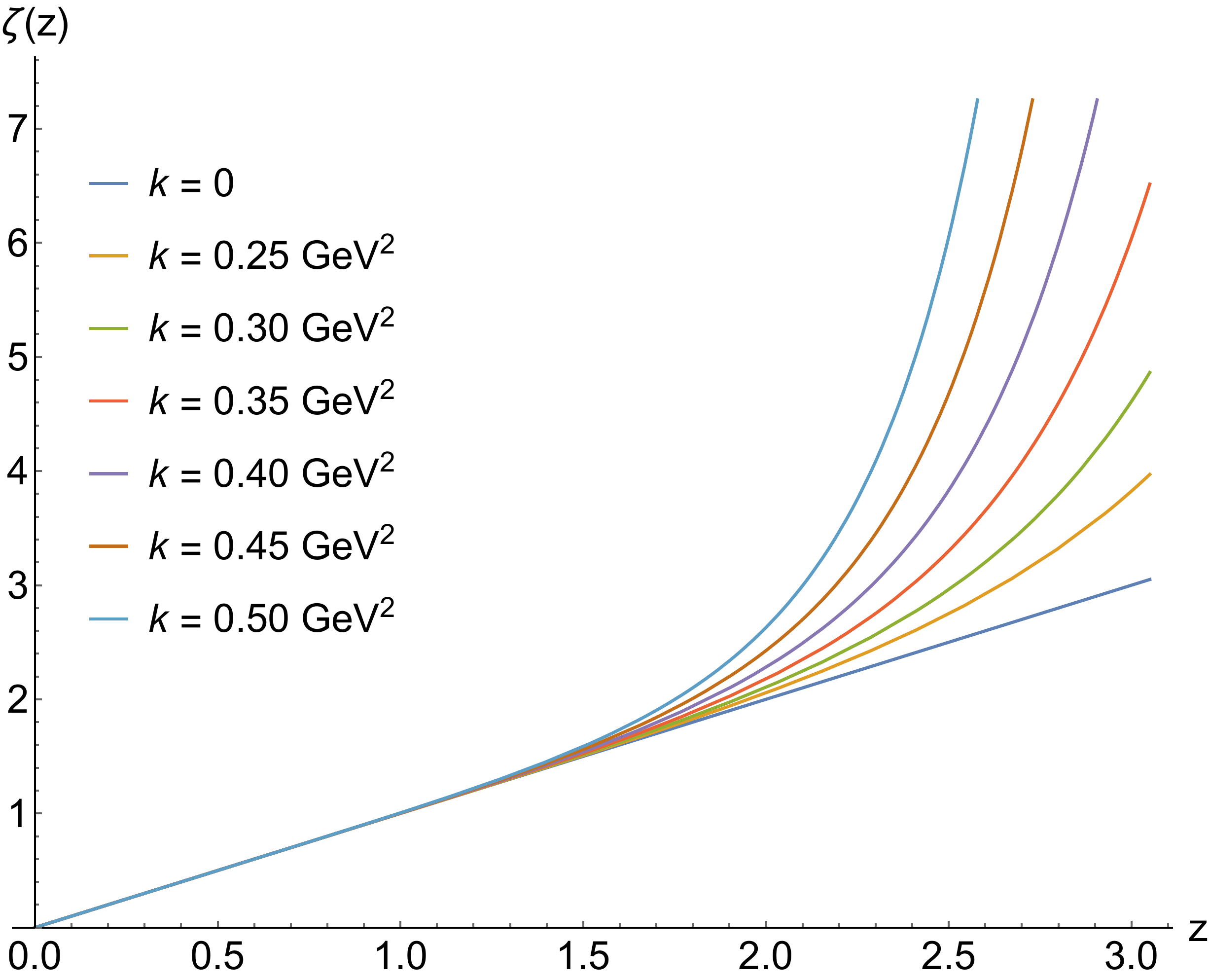}
	\caption{Warp factor $\zeta(z)$, for several values of $k$ in GeV$^2$.}
	\label{fig:zeta}
\end{figure}

\begin{figure}[h]
	\centering
	\includegraphics[scale = 0.26]{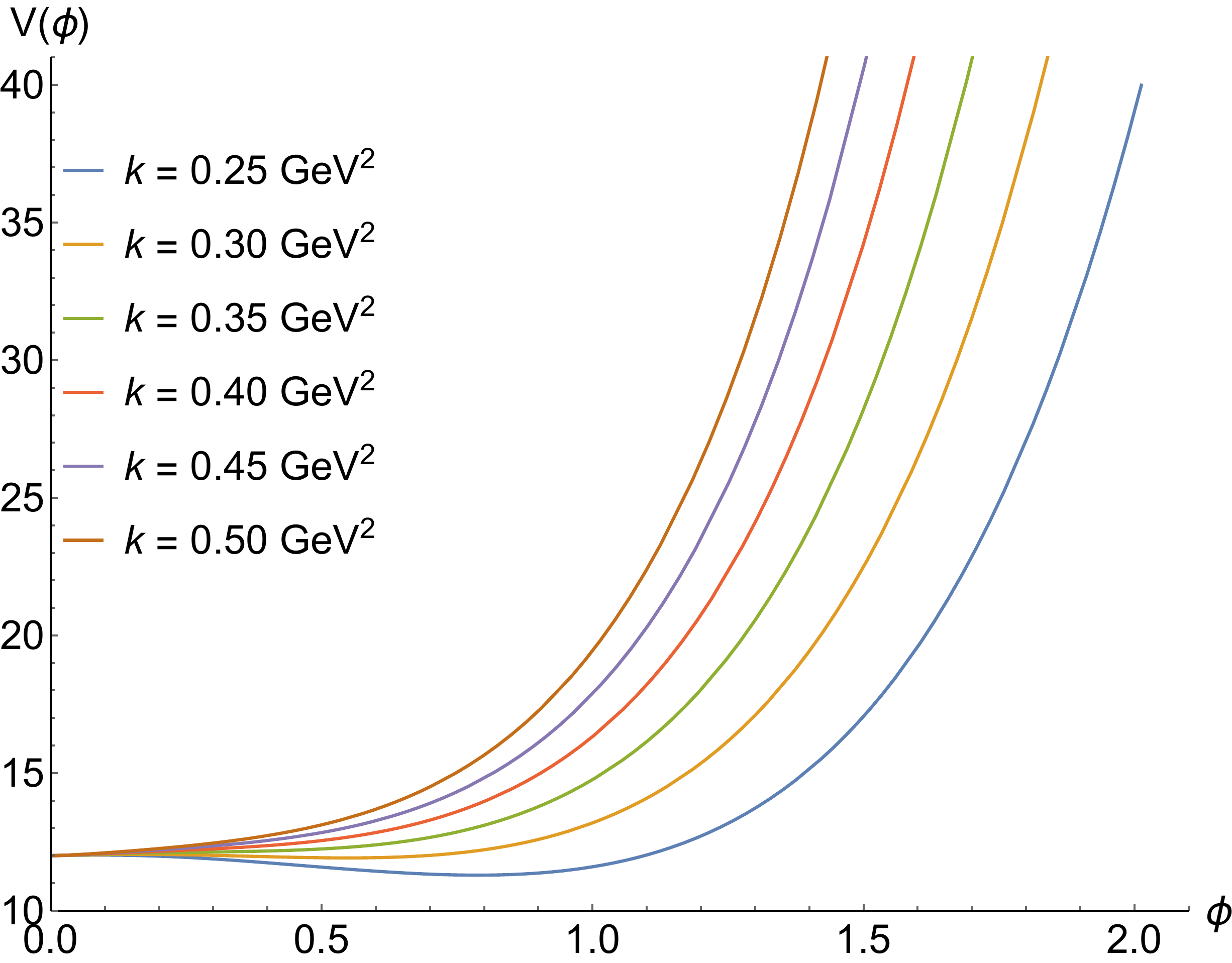}
	\caption{Dilaton potential $V(\phi)$ for several values of $k$ in GeV$^2$.}
	\label{fig:pot_dilaton}
\end{figure}

\begin{figure}[h]
	\centering
	\includegraphics[scale = 0.26]{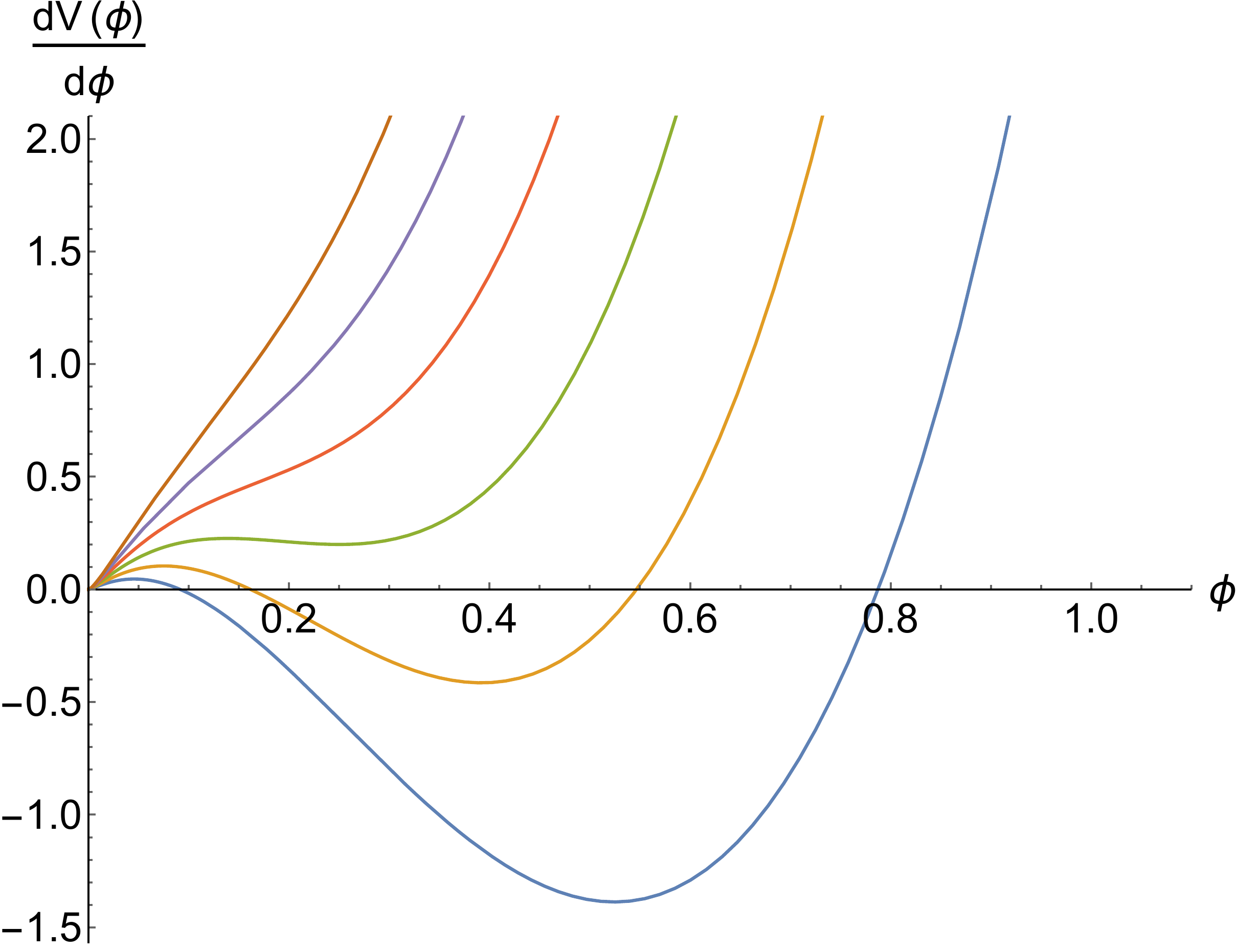}
	\caption{First derivative of the dilaton potential $V(\phi)$ for several values of $k$ in GeV$^2$. The color legends for each value of $k$ in this case are the same as in Fig. \ref{fig:pot_dilaton}.}
	\label{fig:pot_dilatonprime}
\end{figure}

In Fig. \ref{fig:dilaton} the exponential dilaton profile is shown, for several values of $k$ in GeV$^2$.
By using a shooting method, one can numerically solve Eqs. (\ref{breqn3}, \ref{Veqn3}), integrating from the UV boundary towards the IR one. The boundary condition $\zeta(z)\to z$ was used in the UV, i.e., one imposes the background asymptotically approaches AdS, in the UV. The numerical solution for the warp factor $ \zeta(z) $ is displayed in Fig. \ref{fig:zeta}, for several values of $k$, including $k=0$, where one recovers the pure AdS case, i.e, $\zeta(z)\to z$. 
In Figs. \ref{fig:pot_dilaton} and \ref{fig:pot_dilatonprime}, respectively, the dilaton potential $V(\phi)$, given by \eqref{Veqn3}, and its first derivative as a function of $\phi$, are respectively displayed, for several values of $k$.

One can see that there are some values of $ k $ for which the potential is non-monotonic. As discussed in Ref. \cite{Gursoy:2008za}, here we will take a value of $k$ for which the potential is a monotonically increasing function, to prevent the appearance of conformal fixed points, making sure that the AdS is the true vacuum solution for $\phi=0$. For this reason we anticipate that $k = 0.35$ GeV$^2$ will be fixed, for the exponential dilaton profile in Eq. \eqref{expdilaton}. A no larger value for $k$ could be chosen, otherwise, the mass of the scalar glueball and higher-spin resonances would be too high, in comparison with the literature \cite{MarinhoRodrigues:2020yzh}.
.

\subsection{Glueball Action and Mass Spectra}
\label{grall}

\aalt{At the classical level, QCD is a scale-invariant theory, which is broken by quantum fluctuations. The pure gluon part of QCD Lagrangian in 4-dimensions is described by 
\beq
{\cal L}_{\cal G} = -\frac14 G^a_{\mu\nu}(x)G^{\mu\nu,a}(x),
\eeq
where 
$ G^a_{\mu\nu}(x)= \partial_\mu A^a_\nu(x) - \partial_\nu A^a_\mu(x)+ gf^{abc}A^b_\mu(x)A^c_\nu(x)$, for $A_{\mu}(x)$ being the gluon field with $a = 1, \ldots, 8$ color indexes. 
Scalar glueballs states are interpolated by the lowest-dimensional gluonic operator $\mathcal{O}(x)=G^a_{\mu\nu}(x)G^{\mu\nu,a}(x)$, which carries vacuum quantum
numbers. 
The glueball action, in the string frame, reads \cite{Li:2013oda,FolcoCapossoli:2015jnm}
\begin{equation}\label{acao_soft1}
S = \int d^5 x \sqrt{-g} e^{-\phi(z)}\left[g^{MN} \partial_M{\cal G}\partial_N{\cal G} + M^2_{5} {\cal G}^2\right],
\end{equation}
where $ M_5 $ is the mass of the scalar field $ {\cal G}(x) $ in 5 dimensions. Action \eqref{acao_soft1} is associated with the local gauge-invariant QCD boundary operator ${\rm tr}(G_{\mu\nu}G^{\mu\nu})$, whose dual string modes ${\cal G}(x^\mu, z)$ represent normalizable
solutions of the scalar wave equation in the bulk geometry with the UV
behavior $\lim_{z\to0}{\cal G}(x, z)=z^\Delta{\cal G}(x)$, where $\Delta$ is the conformal dimension. One assumes the glueball can be excited from the QCD vacuum and described by either the quenched dynamical holographic model, as introduced in Ref. \cite{Li:2013oda}, yielding Eq. \eqref{acao_soft1}, or in other similar contexts, as in Ref. \cite{FolcoCapossoli:2015jnm}. Eq. \eqref{eom_1} is therefore obtained when one takes the variation of the action \eqref{acao_soft1} with regard to the scalar glueball field
${\cal G}$. This dynamical
softwall model has been used to describe the mass of scalar glueball states and its radial excitations with good agreement with lattice data. }

\aalt{When one takes the variation of the action \eqref{acao_soft1} with regard to the scalar glueball field
${\cal G}$}, the equations of motion from \eqref{acao_soft1} read
\begin{equation}\label{eom_1}
\!\!\!\!\!\partial_M[\sqrt{-g_s} e^{-\phi(z)} g^{MN} \partial_N {\cal G}] \!-\! \sqrt{-g_s} e^{-\phi(z)} {{M}^2_{\rm 5}} {\cal G} \!=\! 0\,. 
\end{equation}
Using the ansatz 
\blt{\begin{eqnarray}
{\cal G}(x^{\mu},z) &=& e^{iq^{\mu}x_{\mu}\!+\frac{B(z)}{2}}\psi(z),\label{stat}
\end{eqnarray}
where 
\begin{eqnarray}
B(z) \!=\! \phi+3\ln{\zeta_{s}(z)},\!\label{stat1}
\end{eqnarray}}
one obtains a Schr\"odinger-like equation given by
\begin{equation}\label{Schrodinger}
-\psi'' + V_{\scalebox{.53}{Sch}}(z)\,\psi = (-q^2)\,\psi,
\end{equation}
where $-q^2=m_n^2$, where $m_n$ are the glueball masses and \blt{$n$ represents the radial excitations, with $n=0$ being the ground state. Note that in this section, and in the next one as well, we are going to consider only the ground state masses}. Finally, $V_{\scalebox{.53}{Sch}}$ is the Schr\"odinger potential given by
\begin{equation}\label{SchrPot}
V_{\scalebox{.53}{Sch}}(z) = \left(\frac{B'^2}{4}-\frac{B''}{2} + M_5^2\,\zeta^{-2}\,e^{\frac{4}{3}\,\phi(z)}\right),
\end{equation}
where the relation \eqref{zetastring} between $\zeta_s$ and $\zeta(z)$ was used. The relation between string frame quantities and Einstein frame ones, already obtained in the previous subsections, is the following
\begin{eqnarray}
\zeta_s(z) &=& \zeta(z)\,e^{-\frac{2}{3}\,\phi(z)}, \label{zetastring}\\
V_s(\phi) &=& V(\phi)\,e^{-\frac{4}{3}\,\phi(z)},
\label{potstring}
\end{eqnarray}
where the subscript ``$s$" denotes a quantity in the string frame, whereas $\zeta(z)$ and $V(\phi)$, respectively given by \eqref{zetasol} and \eqref{dilpotsol}, are in the Einstein frame, for the quadratic dilaton case, and \eqref{zetalinear} and \eqref{potlinear} for the linear dilaton case.

For higher spin fields in AdS, the following mass relation will be considered \cite{Ballon-Bayona:2015wra}
\begin{equation}\label{massa1}
M_5^2 = J(J+4) - J + \gamma(J),
\end{equation}
where one already considers twist-4 $(\Delta-J = 4)$ even-spin glueball operators ${\cal O}_{4+J}\equiv J^{PC}$, with $P=C=+1$, where $ P $ is the parity and $ C $ the charge conjugation. Besides, a contribution coming from the anomalous dimension $\gamma(J)$ is also included, for the glueball operator.

Concerning the anomalous dimensions, we are going to consider it depending on the spin $J$ in two different forms, which we will call by \emph{anomalous I} and \emph{anomalous II}, respectively given by:
\begin{eqnarray}
\gamma^{I}(J) &=& \gamma_0 J \quad {(\rm anomalous\; I)} \label{AnomalousI}\\
\gamma^{II}(J) &=& \gamma_0 \ln{(1+J)} \quad {(\rm anomalous\; II)} \label{AnomalousII}
\end{eqnarray}
so that for $J=0$, $\gamma^{II}(J)$ reduces to $\gamma^{I}(J)$.

\subsubsection{Quadratic Dilaton}

For the quadratic dilaton profile, Eq. \eqref{Schrodinger} was solved numerically in \cite{Li:2013oda, Capossoli:2016kcr, Capossoli:2016ydo} in different contexts. In Ref. \cite{FolcoCapossoli:2016ejd}, Eq. \eqref{Schrodinger} was analytically solved, giving glueball masses in good agreement with QCD lattice data. 

The mass spectra of the \blt{even-spin glueball} family, for both the lattice QCD and the dynamical AdS/QCD model, and taking into account both the anomalous dimensions I and II, given by \eqref{AnomalousI} and \eqref{AnomalousII}, respectively,
are displayed in Tables \ref{CES}. The values of $k$ and $\gamma_0$ were chosen to better fit low $J^{PC}$ \blt{even-spin glueball} resonances with lattice QCD data (for more details see \cite{FolcoCapossoli:2016ejd} and references therein).
\begin{table}[H]
	\begin{center}\medbreak
		\begin{tabular}{||cc||c||c||c||c||c||c||c|}
			\hline\hline
			&  $J^{PC}$ & Mass${}_{\scriptsize{{\rm I}}}$ & Mass${}_{\scriptsize{{\rm II}}}$ & Mass \scriptsize{[latt]} & Mass${}_{\scalebox{.70}{\textsc{I}}}^{\scalebox{.70}{{{\cite{Szanyi:2019kkn}}}}}$&Mass${}_{\scalebox{.70}{\textsc{II}}}^{\scalebox{.70}{{{\cite{Szanyi:2019kkn}}}}}$ \\\hline\hline
			\,&\, 0 \,&\, 1.84 \,&\,0.82\,& 1.595 \, & & \\\hline
			\,&\, 2 \,&\, 2.64 \,&\,2.11\,& 2.39 \, & 1.76&1.758 \\\hline
			\,&\, 4 \,&\, 3.52 \,&\,3.21\,&\, 3.80 \, & 3.19&3.198 \\\hline
			\,&\, 6 \,&\, 4.42 \,&\,4.25\,&\, 4.48 \, & 4.25&4.249 \\\hline
			\,&\, 8 \,&\, 5.31 \,&\,5.26\,&\, \,& 5.16 &5.165\\\hline
			\,&\, 10 \,&\, 6.21 \,&\,6.27\,&\, \,& & \\\hline
			\,&\, 12 \,&\, 7.10 \,&\,7.64\,&\, \,& & \\\hline
			\,&\, 14 \,&\, 7.99 \,&\,8.64\,&\, \,&& \\\hline
			\,&\, 16 \,&\, 8.88 \,&\,9.63\,&\, \,&& \\\hline
			\,&\, 18 \,&\, 9.77 \,&\,10.6\,&\, \,&& \\\hline
			\hline\hline
		\end{tabular}\caption{Mass spectra (GeV) of even-spin glueballs as a function of $J^{PC}$,\cclt{for the quadratic dilaton model 
 in the dynamical AdS/QCD model using the anomalous dimension I \eqref{AnomalousI} (second column) and II \eqref{AnomalousII} (third column), in the lattice QCD (fourth column) \cite{FolcoCapossoli:2016ejd}, and in DP Regge model \cite{Szanyi:2019kkn} (fifth and sixth columns, respectively). }}
		\label{CES}
	\end{center}
\end{table}

The results presented in Table \ref{CES} were taken from Ref. \cite{FolcoCapossoli:2016ejd}, for the anomalous I \cclt{and II cases}. For the anomalous II case, we have just used the analytically mass spectra obtained therein and computed the glueball masses, substituting the anomalous dimension I by II, given by \eqref{AnomalousII}.

\begin{figure}[h]
	\centering
	\includegraphics[width=8cm]{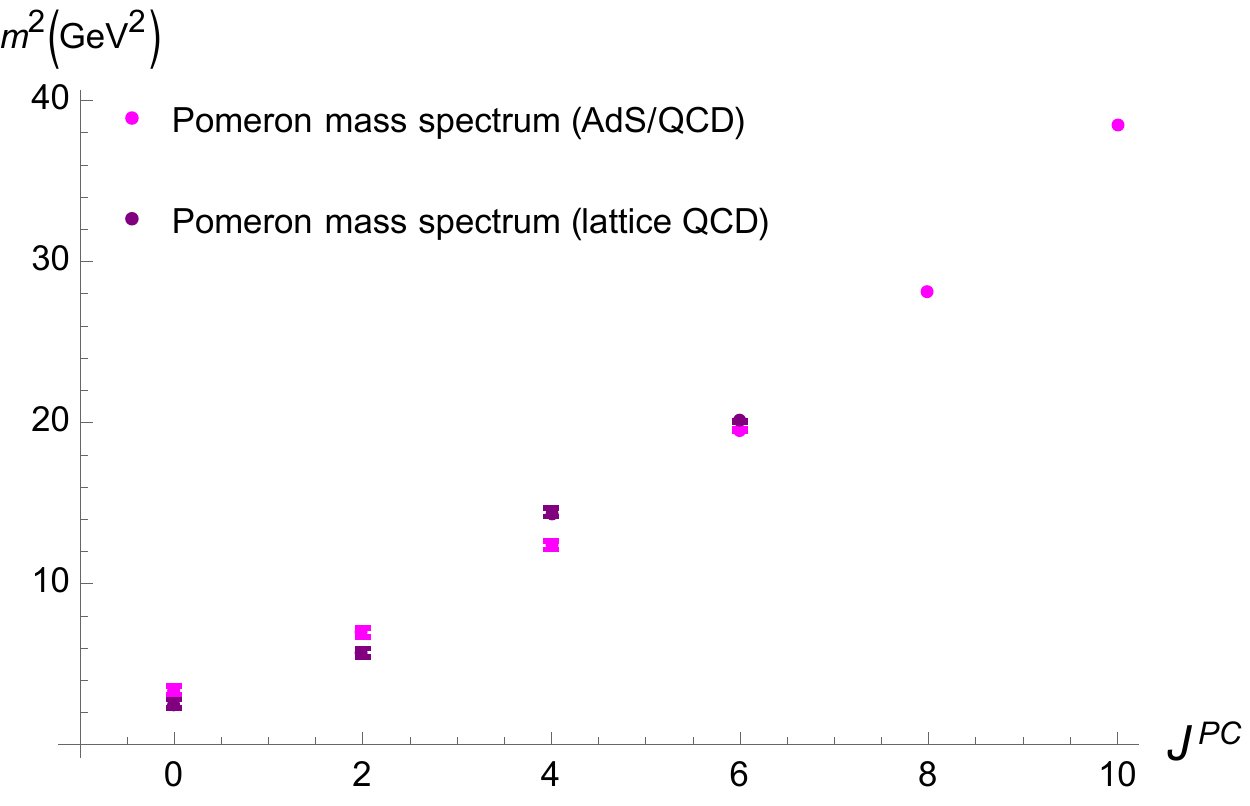}
	\caption{(Squared) mass spectra of even-spin glueballs as a function of $J^{PC}$, in both in lattice QCD (purple points), containing just even-spin glueballs, for $J^{PC}=0,2,4,6$; and in the AdS/QCD model (magenta points), for $J^{PC}=0,2,\ldots, 10$, with the respective error bars.}
	\label{figm2xj}
\end{figure}

As an example, with the data in the second column of Table \ref{CES} one can reproduce the Regge trajectory of the \blt{even-spin glueball} family, illustrated by Fig. \ref{figm2xj} in the dynamical AdS/QCD model. Besides, using data in the third column of Table \ref{CES}, the Regge trajectory of the \blt{even-spin glueball} family, is depicted in Fig. \ref{figm2xj} in lattice QCD. 

\subsubsection{Linear Dilaton}

Now, we turn to the linear dilaton profile. Despite the fact that the linear dilaton is simpler, compared to the quadratic one, we could not find any analytical solution for Eq. \eqref{Schrodinger}, as obtained in \cite{FolcoCapossoli:2016ejd}. In this way, we had to solve it numerically, through the \emph{NDEingensystem} routine in {\tt{Mathematica}}, to obtain the mass spectra from \eqref{Schrodinger}. The best fit parameters were found to be
\begin{eqnarray}
\!\!\!\!\!\!\!\!(\sqrt{k}, \gamma_0) &=& (0.505\;\mathrm{GeV}, -1.995), \quad \mathrm{anomalous\; I}, \\
\!\!\!\!\!\!\!\!(\sqrt{k}, \gamma_0) &=& (0.445\;\mathrm{GeV}, -1.5), \quad \mathrm{anomalous\; II}.
\end{eqnarray}
The results, for the obtained mass spectra, are displayed in Table \ref{CES3}, for both the anomalous I and anomalous II cases, respectively. 

\begin{table}[H]
	\begin{center}\medbreak
		\begin{tabular}{||cc||c||c||c||c||c||c|}
			\hline\hline
				&  $J^{PC}$ & Mass${}_{\scriptsize{{\rm I}}}$ & Mass${}_{\scriptsize{{\rm II}}}$ & Mass \scriptsize{[latt]} & Mass${}_{\scalebox{.70}{\textsc{I/II}}}^{\scalebox{.70}{{{\cite{Szanyi:2019kkn}}}}}$\\\hline\hline
							\,&\, 0 \,&\, 0.88 \,&\,0.82& 1.595 \, & \\\hline
			\,&\, 2 \,&\, 1.90 \,&\,1.95& 2.39 \, & 1.758 \\\hline
			\,&\, 4 \,&\, 3.18 \,&\,3.18& 3.80 \, & 3.198 \\\hline
			\,&\, 6 \,&\, 4.50 \,&\,4.40& 4.48 \, & 4.249 \\\hline
			\,&\, 8 \,&\, 5.83 \,& \,5.62& &5.165\\\hline
			\,&\, 10 \,&\, 7.17 \,&6.83& \,& \\\hline
			\,&\, 12 \,&\, 8.51 \,&8.04& \,& \\\hline
			\,&\, 14 \,&\, 9.86 \,&9.25& \,& \\\hline
			\,&\, 16 \,&\, 11.2 \,&10.4& \,& \\\hline
			\,&\, 18 \,&\, 12.5 \,&11.6& \,& \\\hline
			\hline\hline
		\end{tabular}\caption{Mass spectra (GeV) of even-spin glueballs as a function of $J^{PC}$, \cclt{for the linear dilaton model 
 in the dynamical AdS/QCD model using the anomalous dimension I \eqref{AnomalousI} (second column) and II \eqref{AnomalousII} (third column), in the lattice QCD (fourth column) \cite{FolcoCapossoli:2016ejd}, and in DP Regge model \cite{Szanyi:2019kkn} fifth column).} }		\label{CES3}
	\end{center}
\end{table}

\subsubsection{Exponential Dilaton}

For the exponential dilaton profile \eqref{expdilaton}, we also had to solve numerically Eq. \eqref{Schrodinger}, to obtain the mass spectra. The best fit parameters read 
\begin{eqnarray}
\!\!\!\!\!\!\!\!(k, \gamma_0) &=& (0.35\;\mathrm{GeV^2}, - 5), \quad \mathrm{anomalous\; I}, \\
\!\!\!\!\!\!\!\!(k, \gamma_0) &=& (0.35\;\mathrm{GeV^2}, -9.1), \quad \mathrm{anomalous\; II}.
\end{eqnarray}
The results for the mass spectra obtained are displayed in Table \ref{CES5} for both the anomalous I and II cases. 
\begin{table}[H]
	\begin{center}\medbreak
		\begin{tabular}{||cc||c||c||c||c||c||c|}
			\hline\hline
				&  $J^{PC}$ & Mass${}_{\scriptsize{{\rm I}}}$ & Mass${}_{\scriptsize{{\rm II}}}$ & Mass \scriptsize{[latt]} & Mass${}_{\scalebox{.70}{\textsc{I/II}}}^{\scalebox{.70}{{{\cite{Szanyi:2019kkn}}}}}$\\\hline\hline
			\,&\, 0 \,&\, 2.05 \,&\, 2.05 \,& 1.595 \, & \\\hline
			\,&\, 2 \,&\, 2.05 \,&\, 2.05 \,& 2.39 \, & 1.758 \\\hline
			\,&\, 4 \,&\, 3.37 \,& 3.80&4.00 \, & 3.198 \\\hline
			\,&\, 6 \,&\, 5.00 \,& 4.48&5.93 \, & 4.249 \\\hline
			\,&\, 8 \,&\, 6.69 \,&7.80& \,& 5.165 \\\hline
			\,&\, 10 \,&\, 8.40 \,&8.71& \,& \\\hline
			\,&\, 12 \,&\, 8.80 \,&9.03& \,& \\\hline
			\,&\, 14 \,&\, 9.08 \,&9.28& \,& \\\hline
			\,&\, 16 \,&\, 9.32 \,&9.49& \,& \\\hline
			\,&\, 18 \,&\, 9.51\,&9.67& \,& \\\hline
			\hline\hline
		\end{tabular}\caption{Mass spectra (GeV) of even-spin glueballs as a function of $J^{PC}$, \cclt{for the exponential dilaton model 
 in the dynamical AdS/QCD model using the anomalous dimension I \eqref{AnomalousI} (second column) and II \eqref{AnomalousII} (third column), in the lattice QCD (fourth column) \cite{FolcoCapossoli:2016ejd}, and in DP Regge model \cite{Szanyi:2019kkn} (fifth column).} }
		\label{CES5}
	\end{center}
\end{table}
Interestingly, for the exponential dilaton, in both cases (anomalous I and II) the best fit parameters ($k$ and $\gamma_0$) provided a mass spectrum with degenerate scalar and tensorial glueballs, i.e.,
\begin{equation}
\dfrac{m_{2^{++}}}{m_{0^{++}}}=1.
\end{equation}
Similar findings were first observed in Ref. \cite{Constable:1999gb}, and then in \cite{Ballon-Bayona:2015wra, Rodrigues:2016cdb}, in the context of twist-2 operators, wherein the transversal and traceless part of the spin-2 field $h_{\mu\nu}$ satisfy the same field equation as the scalar field, explaining this degeneracy.

\begin{figure}[h]
	\centering
	\includegraphics[scale = 0.26]{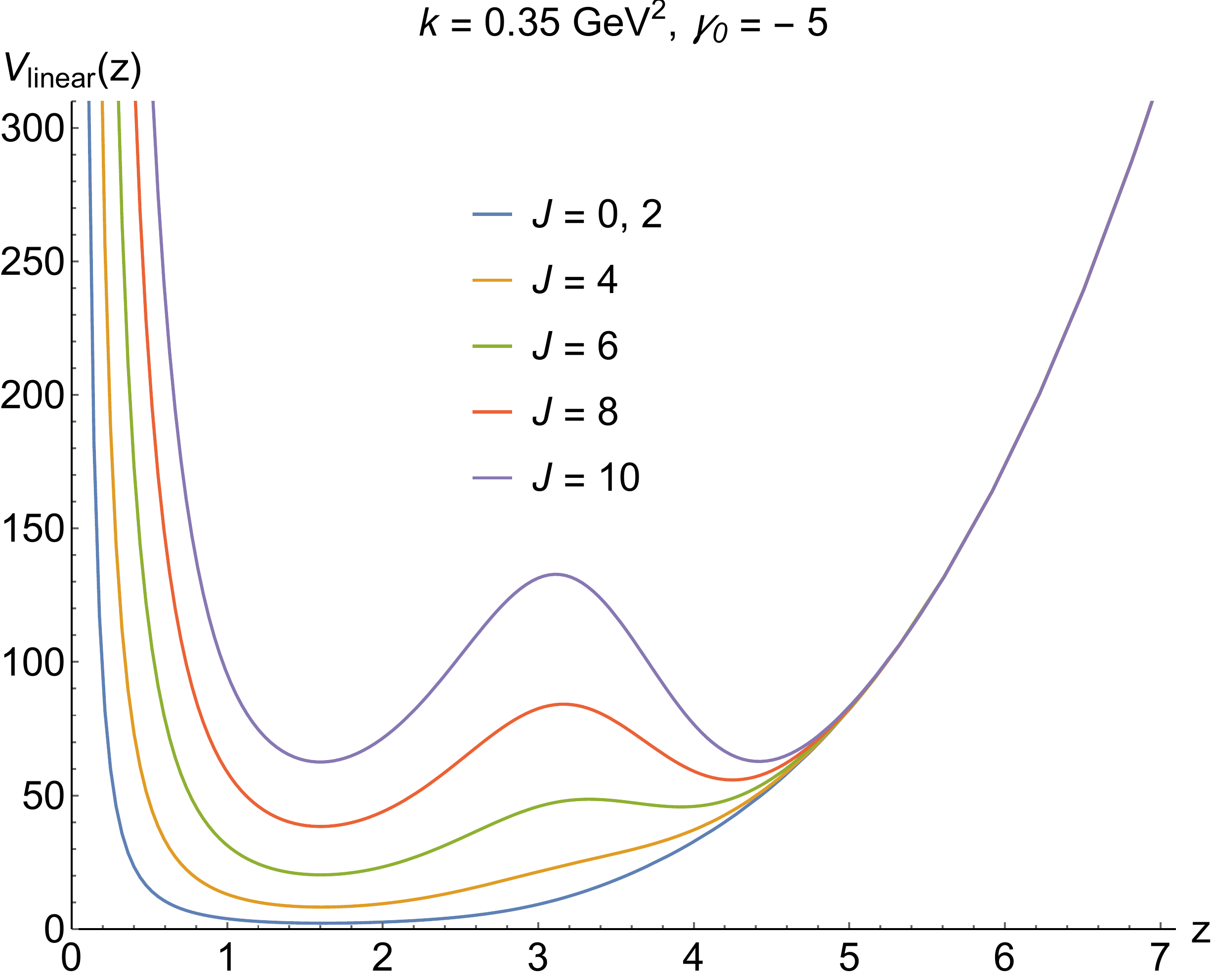}
	\caption{Schr\"odinger potential \eqref{SchrPot} for the anomalous I case, $V_{\rm linear}$, for several values of spin $J$, up to $J=10$.}
	\label{fig:pot_dilatonSchr}
\end{figure}

\begin{figure}[h]
	\centering
	\includegraphics[scale = 0.26]{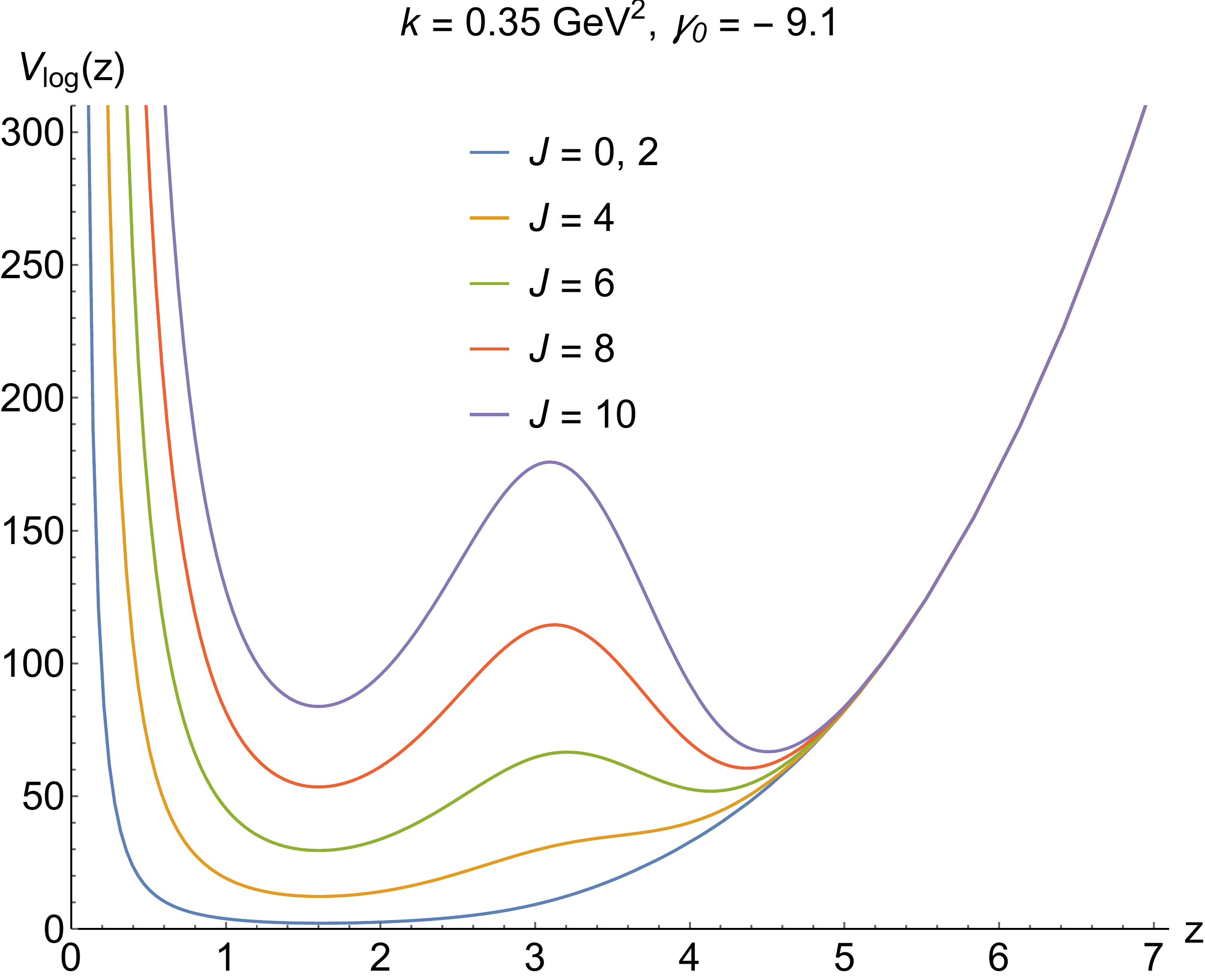}
	\caption{Schr\"odinger potential \eqref{SchrPot} for the anomalous II case, $V_{\rm log}$, for several values of spin $J$, up to $J=10$.}
	\label{fig:pot_dilatonSchr2}
\end{figure}

Finally, for the sake of completeness Figs. \ref{fig:pot_dilatonSchr} and \ref{fig:pot_dilatonSchr2} show the Schr\"odinger potentials \eqref{SchrPot}, for several values of spin $J$ up to $J=10$, for both the anomalous I and II cases, respectively.

\section{CE Regge trajectories of even-spin glueball resonances and mass spectra}\label{ce1}

The fundamental concept underlying the CE regards a measure of 
the correlations among scalar field fluctuations that describe the physical system to be studied. To portray the system, the energy density -- the time component of the stress-energy-momentum tensor $T_{00}(\vec{r})=\uprho(\vec{r})$ -- is the essential ingredient, where $\vec{r}\in\mathbb{R}^n$. The well-known 2-point function $\Upsilon(\vec{r})=\idotsint_{\mathbb{R}^n} \uprho(\vec{\mathfrak{r}}+\vec{r})\uprho(\vec{\mathfrak{r}})\,d\mathfrak{\vec{r}}$ defines a probability distribution \cite{Braga:2018fyc} that makes the CE to be precisely the (Shannon) information entropy of correlations.

For computing the CE, one first calculates the Fourier transform 
\beq\label{fou}
\uprho(\vec{k}) = \frac{1}{(2\pi)^{n/2}}\idotsint_{\mathbb{R}^n}\uprho(\vec{r})e^{-i\vec{k}\cdot \vec{r}}\,d^nx,\eeq 
whose squared norm is the power spectrum. 
 With it in hands, another important quantity, representing the correlation probability distribution in the momentum space, is defined by the so-called modal fraction, as  
\cite{Gleiser:2012tu}
\aalt{\begin{eqnarray}
\mathsf{f}(\vec{k}) = \frac{\left|\uprho(\vec{k})\right|^{2}}{ \idotsint_{\mathbb{R}^n} |\uprho(\vec{k})|^{2}d^n{k}}.\label{modalf}
\end{eqnarray}}
\blt{The modal fraction is the normalized Fourier transform of the 2-point correlation function of the field fluctuations. Therefore, low-power modes are less likely to occur, when compared to higher power modes. } The modal fraction also carries the input of all modes with momentum ${\vec{k}}$ that contribute to the energy density profile. 
The quantity in (\ref{modalf}) measures the contribution of wave modes to the power spectral density associated to the energy density. In fact, standing wave excitation modes can influence all the system components at a fixed frequency associated with these modes. Being the total energy finite, the power spectral density regarding the specific mode $\vec{\mathtt{k}}$, enclosed in the measure $d^n{k}$, is given by $
{P}\left(\vec{\mathtt{k}}\;\vert\; d^n{k}\right)\sim \left|\uprho(\vec{\mathtt{k}})\right|^{2}d^n{k}$ \cite{Gleiser:2018kbq}, referring to the spectral energy distribution found in $d^n{k}$, also measuring how the density spatially fluctuates.
As the power spectrum describes the
fluctuation of the scalar field spatial profile, the modal fraction measures the contribution of a range of modes for the description of the shape of the scalar field. 
Therefore, the differential CE reads 
%The amount of information to describe the spatial profile of $\uprho$, in terms of the Fourier modes, is given by the CE, 
\aalt{\begin{eqnarray}
{\rm CE}_\uprho= - \idotsint_{\mathbb{R}^n}{\mathsf{f}_{\scalebox{.53}{$\natural$}}}(\vec{k})\ln {\mathsf{f}_{\scalebox{.53}{$\natural$}}}(\vec{k})\, d^nk\,,
\label{confige}
\end{eqnarray}
where $\mathsf{f}_{\scalebox{.53}{$\natural$}}(\vec{k})=\mathsf{f}(\vec{k})/\mathsf{f}_{\scalebox{.53}{max}}(\vec{k})$.}

For studying \blt{even-spin glueball} resonances, one can consider $n=1$, corresponding to the $z$ bulk dimension. 
Using the Lagrangian, $L$, of the glueball action \eqref{acao_soft1}, into the stress-energy-momentum tensor  
 \begin{equation}
 \!\!\!\!\!\!\!\!T^{\mu\nu}\!=\! \frac{2}{\sqrt{ -g }}\!\! \left[{\partial_{ g_{\mu\nu}} (\sqrt{-g}{L})}\!-\!\partial_{ x^\alpha } {\partial_{\frac{\partial g_{\mu\nu} }{\partial x^\alpha}} (\sqrt{-g} {L})}
% \!+\!\mathcal{T}^{mn}\!
 \right],
 \label{em1}
 \end{equation} 
 %\noindent  
 and the $\uprho(z)$ energy density, to be used in Eqs. (\ref{fou} -- \ref{confige}), is given by the following expression, \beq\label{t00}
\uprho(z)= T_{00}(z)=\frac{1}{\zeta_s(z)^2}\,\left[({\cal G}'(z))^2 + M_5^2({\cal G}(z))^2\right]\,.
\eeq
Since the \blt{even-spin glueball} family has an energy density associated with, given by Eq. (\ref{t00}), another method for computing the \blt{even-spin glueball} mass spectra, using the CE, can be then implemented. For it, a hybrid model, that takes into account both the lattice QCD model and AdS/QCD, will be employed to derive the mass spectra of higher \blt{even-spin glueball} resonances.

\subsection{Hybridizing lattice QCD data, up to $J^{PC}=6$, and AdS/QCD}
\label{chan1}

\begin{figure}[h]
\centering
\includegraphics[width=8cm]{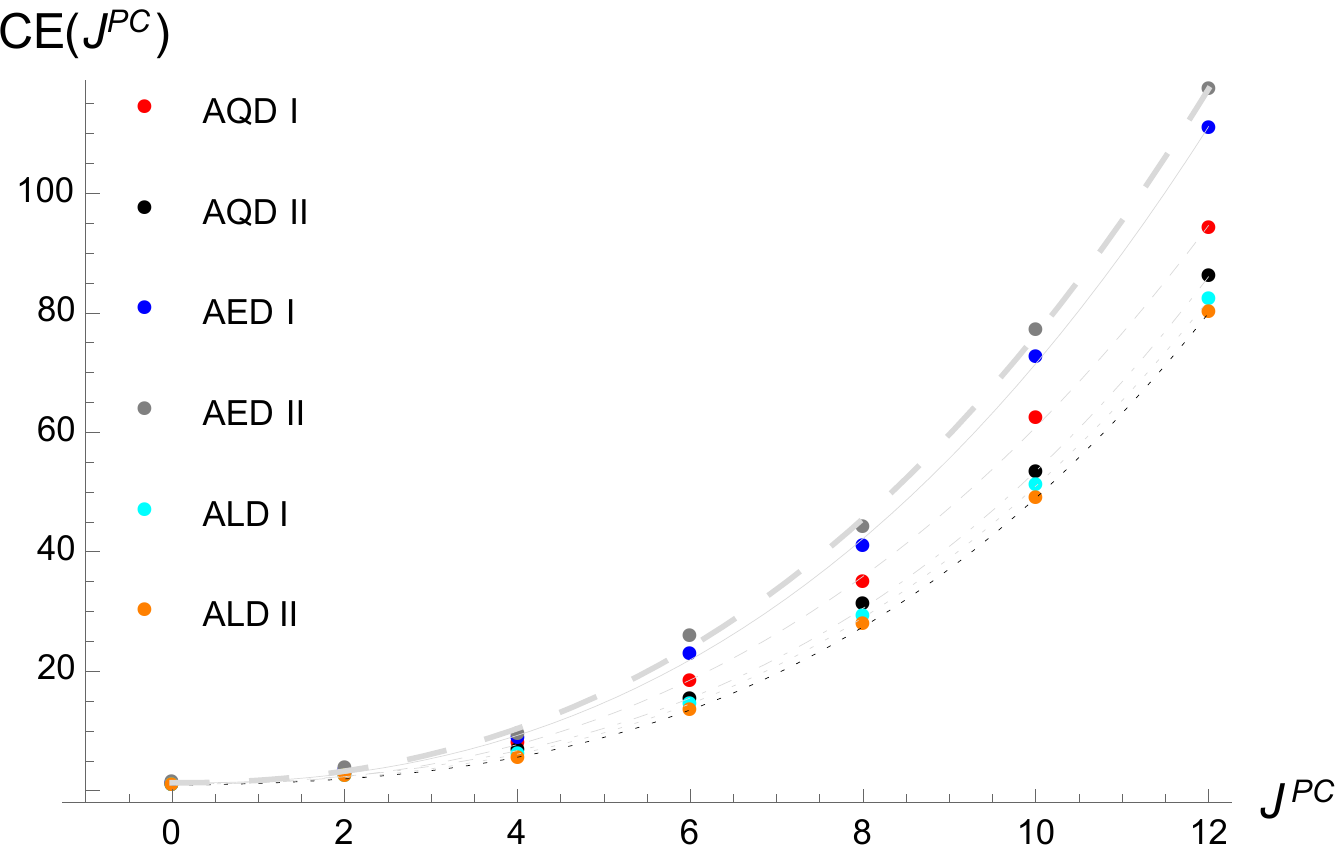}
\caption{CE of the \blt{even-spin glueballs} as a function of $J^{PC}$. The anomalous quadratic [exponential] dilaton I (AQD I) [(AED I)] model is shown by the red [blue] points, whose interpolation continuous curve is given by Eq. (\ref{itp1}) [(\ref{itp1exp})]. The anomalous linear dilaton I (ALD I) [quadratic dilaton II (AQD II)] model is described by cyan [black] points, and inset by the light-gray dotted [dot-dashed light-gray] curve, described by Eq. (\ref{itp1lin}) [(\ref{itp1log})]. The anomalous exponential [linear] dilaton II (AED II) model is illustrated by the gray [orange] points, with large-dashed [black dotted] interpolation curve plot by Eq. (\ref{itp1expii}) [(\ref{itp1linii})].}
\label{figcexj}
\end{figure}

The CE of \blt{even-spin glueballs}, for the six dilaton models, can be numerically computed. As a first analysis, the CE of \blt{even-spin glueballs} can be derived as a function of $J^{PC}$, by Eqs. (\ref{fou} -- \ref{confige}), when one considers lattice QCD data, up to $J^{PC}=6$, and including for $J^{PC}=8, 10$ the \blt{even-spin glueballs} mass spectra (\ref{massa1}). 
The results are compiled in Table \ref{CEJ}. 
\begin{table}[ht]
\begin{center}\medbreak
\begin{tabular}{||c||c||c||c||c||c||c||c||c||}
\hline\hline
  $J^{PC}$ &~CE$_{\scalebox{.53}{AQD I}}$& CE$_{{\scalebox{.53}{AED I}}}$& CE$_{{\scalebox{.53}{ALD I}}}$&~CE$_{\scalebox{.53}{AQD II}}$&~CE$_{{\scalebox{.53}{AED II}}}$&~CE$_{{\scalebox{.53}{ALD II}}}$ \\\hline\hline
   \, 0 \,&$1.00$&1.10&0.88&1.30 &0.92&0.81 \\\hline
   \, 2 \,&$2.98$&3.35&2.43&3.66 &2.62&2.30 \\\hline
   \, 4 \,&$7.82$&8.72&6.02& 9.32&6.62&5.42\\\hline
   \, 6\, &$18.32$&22.72&13.32&25.81 &15.21&14.39 \\\hline
   \, 8\, &$34.78$&40.92&27.72&43.99 &31.08&29.18\\\hline
   \, 10 \,&$62.34$&72.55&48.90&76.94 &53.14&51.01 \\\hline
    \, 12 \,&$94.62$&111.16&79.95&117.46 &86.11&82.12 \\\hline
     \, 14 \,&$138.27$&162.59&121.99&170.59&129.91&124.05 \\\hline
     \, 16 \,&$192.92$&227.17&176.83&236.96 &186.82&178.27\\\hline
      \, 18 \,&$259.71$&306.28&317.90&246.16 &258.46&246.38 \\\hline\hline
\end{tabular}\caption{The CE of \blt{even-spin glueballs} as a function of $J^{PC}$, taking into account lattice QCD data, up to $J^{PC}=6$, and AdS/QCD, for $J^{PC}=8,10$. The second [fifth] column displays the \blt{even-spin glueballs} CE for the anomalous quadratic dilaton I [II] (AQD I [II]) model; the third [sixth] column shows the \blt{even-spin glueballs} CE computed from the anomalous exponential dilaton I [II] (AED I [II]) model; the fourth [seventh] column illustrates the CE of the \blt{even-spin glueball} family in the anomalous linear dilaton I [II] (ALD I [II]) model. }
\label{CEJ}
\end{center}
\end{table}
Data in Table \ref{CEJ} can be then interpolated, to obtain the first type of configurational-entropic (CE) Regge trajectory, 
relating the CE of \blt{even-spin glueballs} resonances to their $J^{PC}$ spin.
Firstly, the expression of the CE Regge trajectory, for the anomalous quadratic dilaton I model, reads
\begin{eqnarray}
 {\rm CE}_{\scalebox{.55}{AQD I}}(J^{PC}) &=& 0.02329\, \left({J^{PC}}\right)^3 +0.39750\, \left({J^{PC}}\right)^2 \nonumber\\&&-0.34629\, {J^{PC}}+ 1.27524.\label{itp1}
  \end{eqnarray} It corresponds to the dashed interpolating curve in Fig. \ref{figcexj}. Cubic interpolation suffices to delimit accuracy within
  $0.42\%$. 
Besides, a similar CE Regge trajectory can be derived, in the anomalous exponential dilaton I  model, representing the continuous interpolating curve in Fig. \ref{figcexj}, as 
\begin{eqnarray}
 {\rm CE}_{\scalebox{.55}{AED I}}(J^{PC}) &=& 0.028715\, \left({J^{PC}}\right)^3 +0.43693\, \left({J^{PC}}\right)^2 \nonumber\\&&-0.22849\, {J^{PC}}+ 1.36214,\label{itp1exp}
  \end{eqnarray}
  \noindent within $0.37\%$ of accuracy. 
    Analogously, the light-gray dotted CE Regge trajectory, for the anomalous linear dilaton I model, interpolates the cyan points in Fig. \ref{figcexj}, and is given by  
\begin{eqnarray}
 {\rm CE}_{\scalebox{.55}{ALD I}}(J^{PC}) &=& 0.0328785\, \left({J^{PC}}\right)^3 +0.15853\, \left({J^{PC}}\right)^2 \nonumber\\&&+0.12701\, {J^{PC}}+ 0.98093,\label{itp1lin}
  \end{eqnarray} with accuracy within $0.23\%$. 
  
  The expression of the CE Regge trajectory, for the anomalous quadratic dilaton II model, interpolates the black points in Fig. \ref{figcexj}, and yields 
\begin{eqnarray}
 {\rm CE}_{\scalebox{.55}{AQD II}}(J^{PC}) &=& 0.0354513\, \left({J^{PC}}\right)^3 +0.13899\, \left({J^{PC}}\right)^2 \nonumber\\&&+0.31736\, {J^{PC}}+ 0.95238,\label{itp1log}
  \end{eqnarray} having accuracy within $0.18\%$. 
  Moreover, the CE Regge trajectory in the anomalous exponential dilaton II  model, is represented by the large-dashed light-gray curve that interpolates gray points, in Fig. \ref{figcexj}, 
\begin{eqnarray}
 {\rm CE}_{\scalebox{.55}{AED II}}(J^{PC}) &=& 0.028021\, \left({J^{PC}}\right)^3 +0.47771\, \left({J^{PC}}\right)^2 \nonumber\\&&-0.09083\, {J^{PC}}+ 1.34452,\label{itp1expii}
  \end{eqnarray}
  \noindent within $1.83\%$ of accuracy. 
   The black dotted CE Regge trajectory for the anomalous linear dilaton II model interpolates the orange points in Fig. \ref{figcexj}, reading  
\begin{eqnarray}
 {\rm CE}_{\scalebox{.55}{ALD II}}(J^{PC}) &=&0.035483\, \left({J^{PC}}\right)^3 +0.10871\, \left({J^{PC}}\right)^2 \nonumber\\&&+0.17159\, {J^{PC}}+ 0.91878.\label{itp1linii}
  \end{eqnarray} Cubic interpolation delimits accuracy within $\sim0.2\%$.

Both data in Table \ref{CEJ} and the first type of CE Regge trajectories (\ref{itp1} -- \ref{itp1linii}) are together displayed in Fig. \ref{figcexj}.

\noindent Higher $J^{PC}$ \blt{even-spin glueballs} resonances can, thus, have their CE extrapolated from the CE Regge trajectories (\ref{itp1} -- \ref{itp1linii}), respectively for each dilaton model.

Among the six dilaton models, Fig. \ref{figcexj} shows that the anomalous exponential dilaton II  model represents more configurationally unstable states, whereas the anomalous linear dilaton II model regards \blt{even-spin glueballs} resonances that are more stable, from the configurational point of view.

For deriving the \blt{even-spin glueballs} mass spectra in the six dilatonic models, we will first use, for the \blt{even-spin glueballs} mass spectra up to $J^{PC}=6$, the third column of Table \ref{CES}. In fact, 
it encompasses an average derivation from lattice 
QCD, including both a finite range of $N_c$ and $N_c\to\infty$. 
Therefore, we use just the available data up to $J^{PC}=6$, from QCD lattice setups, completing for $J^{PC}=8,10$ the \blt{even-spin glueballs} mass spectra obtained from AdS/QCD \eqref{massa1}.
This is displayed in the second column of Table \ref{CES}. Therefore, 
the interpolation fitting of the point, representing the CE as a function of the \blt{even-spin glueballs} mass spectra, will provide the mass 
spectra of the next generation of \blt{even-spin glueballs} resonances, with $J^{PC}=12, 14, 16, 18,\ldots$. 

\begin{figure}[h]
\centering
\includegraphics[width=8.6cm]{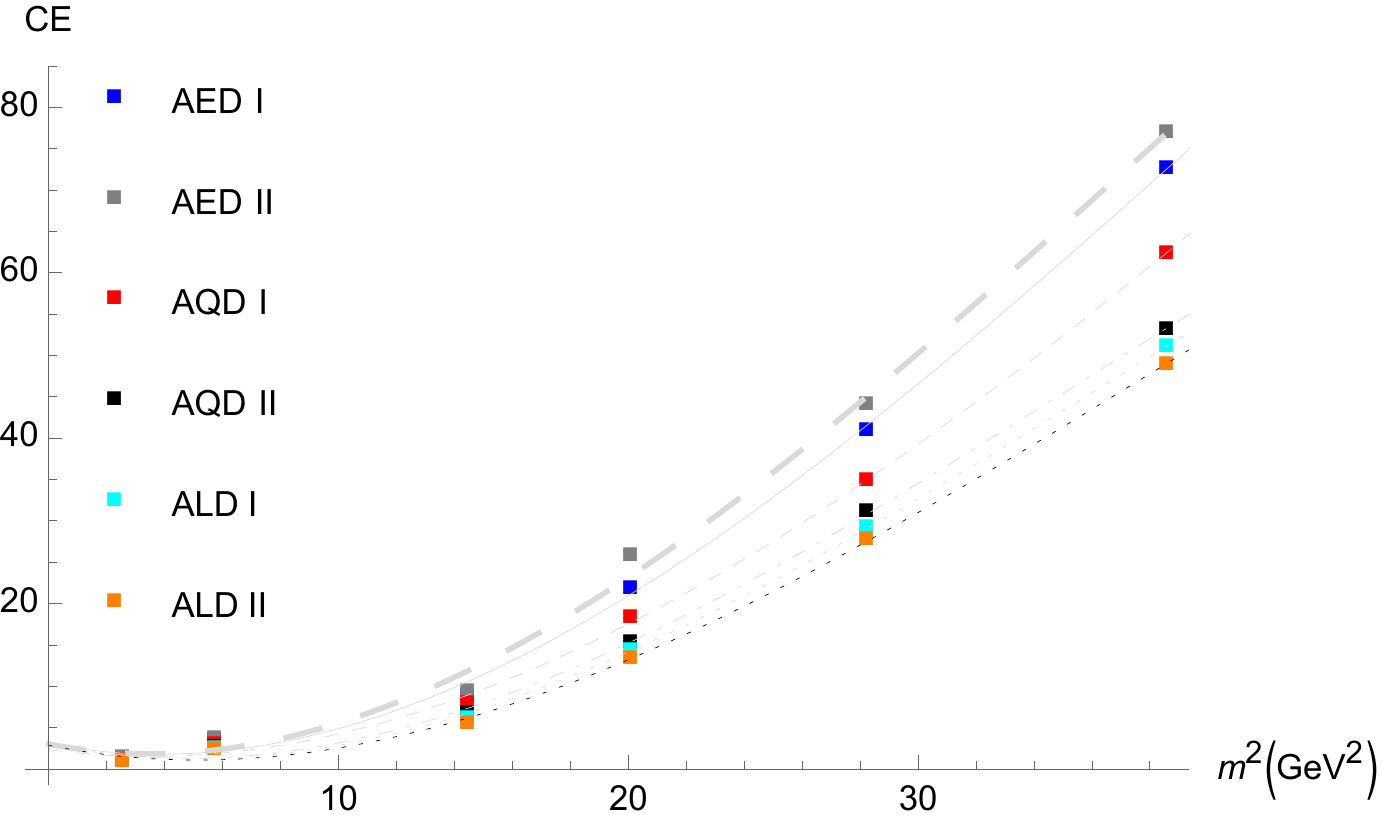}\medbreak
\caption{CE of the \blt{even-spin glueball} family, with respect to the squared mass spectra, taking into account exclusively lattice QCD data, up to $J^{PC}=6$, and including, for $J^{PC}=8, 10$, the \blt{even-spin glueballs} mass spectra (\ref{massa1}). The anomalous quadratic dilaton I [II] model is shown by the red [black] boxes, whereas the anomalous exponential dilaton I [II] model is illustrated by the blue [gray] boxes; the anomalous linear dilaton I [II] model is depicted by cyan [orange] boxes. }
\label{figcexm2}
\end{figure}

The CE Regge trajectory, relating the CE to the \blt{even-spin glueballs} (squared) mass spectra, is shown in Fig. \ref{figcexm2}.
 \noindent Higher $J^{PC}$ \blt{even-spin glueballs} resonances will have their CE extrapolated from the CE Regge trajectories (\ref{itp1}, \ref{itq1}) and (\ref{itp1log}, \ref{itq1log}), for the anomalous quadratic dilaton I and II models, respectively; from the CE Regge trajectories (\ref{itp1exp}, \ref{itq1exp}) and (\ref{itp1expii}, \ref{itq1expii}), respectively for the anomalous exponential dilaton I and II  models; and from Eqs. (\ref{itp1lin}, \ref{itq1lin}) and (\ref{itp1linii}, \ref{itq1linii}), respectively, for the anomalous linear dilaton I and II  models.

The dashed CE Regge trajectory in Fig. \ref{figcexm2} corresponds to Eq. (\ref{itq1}), that interpolates the mass spectra of \blt{even-spin glueballs} for the anomalous quadratic dilaton I model, 
\begin{eqnarray}\label{itq1}
\!\!\!\!\!\!\!\! {\rm CE}_{\scalebox{.55}{AQD I}}(m) \!&=\!&\! -0.0004947 m^6\!+\!0.0719030 m^4\!\nonumber\\&&-0.4808326\,m^2+2.3348292,   \end{eqnarray} within $0.25\%$ standard deviation. This hybrid model, taking into account lattice QCD data up $J^{PC}=6$ and including, for $J^{PC}=8, 10$, the \blt{even-spin glueballs} mass spectra (\ref{massa1}), yields the mass spectra of the $\mathcal{P}_{12}, \mathcal{P}_{14}, \mathcal{P}_{16}$ and $\mathcal{P}_{18}$ \blt{even-spin glueballs} resonances, by employing Eqs. (\ref{itp1}, \ref{itq1}). In fact, for $J^{PC}=12$, Eq. (\ref{itp1}) yields 
${\rm CE} = 94.6198$. Then substituting this value in the CE Regge trajectory (\ref{itq1}), and solving the resulting cubic algebraic equation, the solution is the mass of the $\mathcal{P}_{12}$ \blt{even-spin glueball}, given by 
\beq\label{m12quad}
m_{\mathcal{P}_{12}}=7.0334\; {\rm GeV}.
\eeq Similarly for the next \blt{even-spin glueball} resonance $\mathcal{P}_{14}$, substituting $J^{PC}=14$ into Eq. (\ref{itp1}) implies that the corresponding CE equals to $138.269$. Hence, replacing this value into the CE Regge trajectory (\ref{itq1}) yields 
\beq
m_{\mathcal{P}_{14}}=8.0445\; {\rm GeV}.
\eeq Besides, the same protocol can be accomplished for the $\mathcal{P}_{16}$ \blt{even-spin glueball}, when Eq. (\ref{itp1}) implies that ${\rm CE}_{\mathcal{P}_{16}} = 192.926$. When reinstated in Eq. (\ref{itq1}), it produces the mass, for the ${\mathcal{P}_{16}}$ \blt{even-spin glueball}, 
\beq
m_{\mathcal{P}_{16}}=9.7421\;{\rm GeV}.\eeq
Analogously, for the ${\mathcal{P}_{18}}$, we compute ${\rm CE}_{\mathcal{P}_{18}} = 259.712$, and reinstating into Eq. (\ref{itq1}), it produces the ${\mathcal{P}_{18}}$ \blt{even-spin glueball} mass
\beq
m_{\mathcal{P}_{18}}=10.1372\; {\rm GeV.} 
\eeq

Besides, the continuous curve in Fig. \ref{figcexm2}, that interpolates the blue boxes that fit the mass spectra of \blt{even-spin glueballs}, when the anomalous exponential dilaton I model is employed, is given by  
 \begin{eqnarray}\label{itq1exp}
\!\!\!\!\!\!\!\! {\rm CE}_{\scalebox{.55}{AED I}}(m) \!&=\!&\! -0.00079069 m^6\!+\!0.09515 m^4\!\nonumber\\&&-0.69436\,m^2+3.08280,    \end{eqnarray} within $0.98\%$ standard deviation. 
%This consists of a second type of CE Regge trajectory, this time relating the CE to the pomerons (squared) mass spectra.
For $J^{PC}=12$, Eq. (\ref{itp1exp}) yields 
${\rm CE}_{\mathcal{P}_{12}} = 111.1588$. Then substituting this value into the CE Regge trajectory (\ref{itq1exp}), solving the resulting equation yields the solution
\beq
m_{\mathcal{P}_{12}}=7.1628\; {\rm GeV},
\eeq for the $\mathcal{P}_{12}$ \blt{even-spin glueball} mass. Similarly, for the \blt{even-spin glueball} $\mathcal{P}_{14}$, substituting $J^{PC}=14$ into Eq. (\ref{itp1exp}) implies that CE$_{\mathcal{P}_{14}}=162.597$. Hence, replacing it into 
the CE Regge trajectory (\ref{itq1exp}) yields 
\beq
m_{\mathcal{P}_{14}}=8.8109\; {\rm GeV}.
\eeq Employing a similar procedure for the $\mathcal{P}_{16}$ \blt{even-spin glueball}, Eq. (\ref{itp1exp}) for $J^{PC}=16$ yields ${\rm CE}_{\mathcal{P}_{16}} = 227.179$. When replaced into Eq. (\ref{itq1exp}), it produces the mass, for the ${\mathcal{P}_{16}}$ \blt{even-spin glueball}, \beq
m_{\mathcal{P}_{16}}=9.2062\;{\rm GeV},\eeq and respectively for the ${\mathcal{P}_{18}}$ element, ${\rm CE}_{\mathcal{P}_{18}} = 306.284$, and reinstating into Eq. (\ref{itq1exp}), it produces the ${\mathcal{P}_{18}}$ \blt{even-spin glueball} mass
\beq
m_{\mathcal{P}_{18}}=9.5809\; {\rm GeV.} 
\eeq

Now we analyze the anomalous linear dilaton I model, whose CE Regge trajectory is given by  
 \begin{eqnarray}\label{itq1lin}
\!\!\!\!\!\!\!\! {\rm CE}_{\scalebox{.55}{ALD I}}(m) \!&=\!&\! -0.000708092 m^6\!+\!0.078713 m^4\!\nonumber\\&&-0.73749\,m^2 +3.05330,  \end{eqnarray} within $0.83\%$ standard deviation.
When one fixes $J^{PC}=12$, Eq. (\ref{itp1lin}) yields 
${\rm CE} = 82.117$. Therefore, this value is reinstated in the CE Regge trajectory (\ref{itq1lin}), and solving the resulting equation, the solution is the mass 
\beq
m_{\mathcal{P}_{12}}=7.3861\; {\rm GeV,}
\eeq for the $\mathcal{P}_{12}$ \blt{even-spin glueball}. It yields the range $7.35\; {\rm GeV}\lesssim m_{\mathcal{P}_{12}}\lesssim 7.42\;{\rm GeV}$. Similarly for the next \blt{even-spin glueball} resonance, $\mathcal{P}_{14}$, substituting $J^{PC}=14$ into Eq. (\ref{itp1log}) implies that the CE equals $124.049$. Replacing it into 
the CE Regge trajectory (\ref{itq1lin}) yields \beq
m_{\mathcal{P}_{14}}=8.6039\; {\rm GeV,}\eeq
 being the range $8.58 \,{\rm GeV}\lesssim m_{\mathcal{P}_{14}}\lesssim 8.62\, {\rm GeV}$ trustworthy. A similar procedure is used for the $\mathcal{P}_{16}$ \blt{even-spin glueball}, when Eq. (\ref{itp1lin}) implies that ${\rm CE}_{\mathcal{P}_{16}} = 178.267$, for $J^{PC}=16$. When reinstated in Eq. (\ref{itq1lin}), it yields the mass 
 \beq
 m_{\mathcal{P}_{16}}=9.0158\; {\rm GeV}.
 \eeq Now, replacing $J^{PC}=18$ in Eq. (\ref{itp1lin}) implies that the corresponding CE has value $246.378$. Therefore, solving for it, Eq. (\ref{itq1log}) yields
 \beq\label{m18log2}
 m_{\mathcal{P}_{18}}=9.4091\; {\rm GeV}. 
\eeq

The CE Regge trajectory interpolating curve, corresponding to the anomalous quadratic dilaton II model reads 
 \begin{eqnarray}\label{itq1log}
\!\!\!\!\!\!\!\! {\rm CE}_{\scalebox{.55}{AQD II}}(m) \!&=\!&\! -0.00078764 m^6\!+\!0.083448 m^4\!\nonumber\\&&-0.74445\,m^2 +3.01507,  \end{eqnarray} within $0.79\%$ standard deviation.
The mass spectra for the $\mathcal{P}_{12}, \mathcal{P}_{14}, \mathcal{P}_{16}$ and $\mathcal{P}_{18}$ \blt{even-spin glueballs} resonances can be straightforwardly deduced, when one uses Eqs. (\ref{itp1log}, \ref{itq1log}). In fact, for $J^{PC}=12$, Eq. (\ref{itp1log}) yields 
${\rm CE} = 86.035$. Then substituting this value in (\ref{itq1log}), and solving the resulting equation, the solution is the mass 
\beq
m_{\mathcal{P}_{12}}=7.5618\; {\rm GeV}.
\eeq It yields the range $7.54\; {\rm GeV}\lesssim m_{\mathcal{P}_{12}}\lesssim 7.58\;{\rm GeV}$. Similarly for the next \blt{even-spin glueball} resonance $\mathcal{P}_{14}$, substituting $J^{PC}=14$ into Eq. (\ref{itp1log}) implies the CE equals $129.916$. Hence, replacing this value into 
the CE Regge trajectory (\ref{itq1log}) yields \beq
m_{\mathcal{P}_{14}}=8.4893\; {\rm GeV,}\eeq
 being the mass range $8.51 \,{\rm GeV}\lesssim m_{\mathcal{P}_{14}}\lesssim 8.47\, {\rm GeV}$ a sound one. A similar procedure is used for the $\mathcal{P}_{16}$ \blt{even-spin glueball}, when Eq. (\ref{itp1log}) implies that ${\rm CE}_{\mathcal{P}_{16}} = 186.8200$. When reinstated in Eq. (\ref{itq1log}), it yields the mass 
 \beq
 m_{\mathcal{P}_{16}}=8.8991\; {\rm GeV}.
 \eeq Now, replacing $J^{PC}=18$ in Eq. (\ref{itp1log}) implies that the corresponding CE has value $258.4495$. Therefore, solving Eq.(\ref{itq1log}) with this value yields
 \beq\label{m18log4}
 m_{\mathcal{P}_{18}}=9.2907\; {\rm GeV}. 
\eeq

The CE Regge trajectory interpolating curve, corresponding to the anomalous exponential dilaton II model reads 
 \begin{eqnarray}\label{itq1expii}
\!\!\!\!\!\!\!\! {\rm CE}_{\scalebox{.55}{AED II}}(m) \!&=\!&\! -0.000929376\!+\!0.103323 m^4\!\nonumber\\&&-0.688707\,m^2 +3.01135,  \end{eqnarray} within $0.89\%$ standard deviation.
Now, for $J^{PC}=12$, Eq. (\ref{itp1expii}) yields 
${\rm CE} = 117.46$. Therefore, when one replaces it into (\ref{itq1expii}), it yields mass 
\beq
m_{\mathcal{P}_{12}}=7.2542\; {\rm GeV}.
\eeq In a similar way, for the next \blt{even-spin glueball} resonance $\mathcal{P}_{14}$, substituting $J^{PC}=14$ into Eq. (\ref{itp1expii}) implies that the CE has value $170.593$. Hence, replacing this value into 
 (\ref{itq1}) yields \beq
m_{\mathcal{P}_{14}}=8.60425\; {\rm GeV}.\eeq
 An analogous procedure is used for the $\mathcal{P}_{16}$ \blt{even-spin glueball}, when Eq. (\ref{itp1expii}) implies that ${\rm CE}_{\mathcal{P}_{16}} = 236.957$. When reinstated in Eq. (\ref{itq1expii}), it yields the mass 
 \beq
 m_{\mathcal{P}_{16}}=8.98653\; {\rm GeV}.
 \eeq Now, replacing $J^{PC}=18$ in Eq. (\ref{itp1expii}) implies that the CE has value $317.905$. Therefore, solving for it Eq. (\ref{itq1expii}) yields
 \beq\label{m18log5}
 m_{\mathcal{P}_{18}}=9.3489\; {\rm GeV}. 
\eeq

Now, the last model to be analyzed is the anomalous linear dilaton II model, whose CE Regge trajectory is given by  
 \begin{eqnarray}\label{itq1linii}
\!\!\!\!\!\!\!\! {\rm CE}_{\scalebox{.55}{ALD II}}(m) \!&=\!&\! -0.00064434 m^6\!+\!0.0743293 m^4\!\nonumber\\&&-0.71216\,m^2 +2.85792,  \end{eqnarray} within $0.83\%$ standard deviation.
When one fixes $J^{PC}=12$, Eq. (\ref{itp1linii}) yields 
${\rm CE} = 79.945 $. Therefore, Eq. (\ref{itq1linii}) yields 
\beq
m_{\mathcal{P}_{12}}=7.3589\; {\rm GeV.}
\eeq Similarly for the next member $\mathcal{P}_{14}$ of the \blt{even-spin glueball} family, substituting $J^{PC}=14$ into Eq. (\ref{itp1linii}) implies that the CE equals $121.990$. Consequently, replacing this value into 
 (\ref{itq1linii}) yields \beq
m_{\mathcal{P}_{14}}=8.7266\; {\rm GeV}.\eeq
 A similar protocol is utilized for the $\mathcal{P}_{16}$ \blt{even-spin glueball}, when Eq. (\ref{itp1linii}) implies that ${\rm CE}_{\mathcal{P}_{16}} = 176.836$. When reinstated in Eq. (\ref{itq1linii}), it yields the mass 
 \beq
 m_{\mathcal{P}_{16}}=9.1537\; {\rm GeV}.
 \eeq Now, replacing $J^{PC}=18$ in Eq. (\ref{itp1linii}) implies that the corresponding CE has value $246.161$. Therefore, solving for it Eq. (\ref{itq1linii}) yields
 \beq\label{m18log6}
 m_{\mathcal{P}_{18}}=9.5615\; {\rm GeV}. 
\eeq

The mass spectra of \blt{even-spin glueballs} as a function of the \blt{even-spin glueball} mass $m$ (GeV), taking into account exclusively lattice QCD data up to $J^{PC}=6$, and including for $J^{PC}=8, 10$ the \blt{even-spin glueballs} mass spectra (\ref{massa1}) of AdS/QCD, in the six dilaton models, are summarized in Table \ref{CEM2}.
\begin{table}[H]
\begin{center}\medbreak
\begin{tabular}{||c||c||c||c||c||c||c||c||c||}
\hline\hline
   $J^{PC}$ & $m_{\scalebox{.55}{AQD I}}$&  $m_{\scalebox{.55}{AED I}}$&$m_{\scalebox{.55}{ALD I}}$& $m_{\scalebox{.55}{AQD II}}$&$m_{\scalebox{.55}{AED II}}$&$m_{\scalebox{.55}{ALD II}}$ \\\hline\hline
    \, 12 \,&7.03&7.16&7.38&7.56&7.25&7.35 \\\hline
     \, 14 \,&8.04&8.81&8.60&8.48&8.60&8.72 \\\hline
      \, 16 \,&9.74&9.21&9.01&8.90&8.98&9.15\\\hline
      \, 18 \,&10.14&9.58&9.40&9.29&9.34&9.56 \\\hline\hline
\end{tabular}\caption{Mass spectra (GeV) of \blt{even-spin glueballs} as a function of $J^{PC}$, taking into account exclusively lattice QCD data up to $J^{PC}=6$, and including for $J^{PC}=8, 10$ the \blt{even-spin glueballs} mass spectra (\ref{massa1}). The second and fifth columns display the CE of the anomalous quadratic dilaton I and II (AQD I and II) models; the third and sixth columns show the \blt{even-spin glueballs} CE computed from the anomalous exponential dilaton I and II (AED I and II) models; the fourth and seventh columns depict the CE computed from the anomalous linear dilaton I and II (ALD I and II) models. }
\label{CEM2}
\end{center}
\end{table}

\subsection{Using just lattice QCD data up to $J^{PC}=6$}
\label{chan2}

Now, using exclusively the available data up to $J^{PC}=6$, from lattice QCD data,
the CE of the \blt{even-spin glueball} family can be computed by Eqs. (\ref{fou} -- \ref{confige}).
\begin{figure}[h]
\centering
\includegraphics[width=8cm]{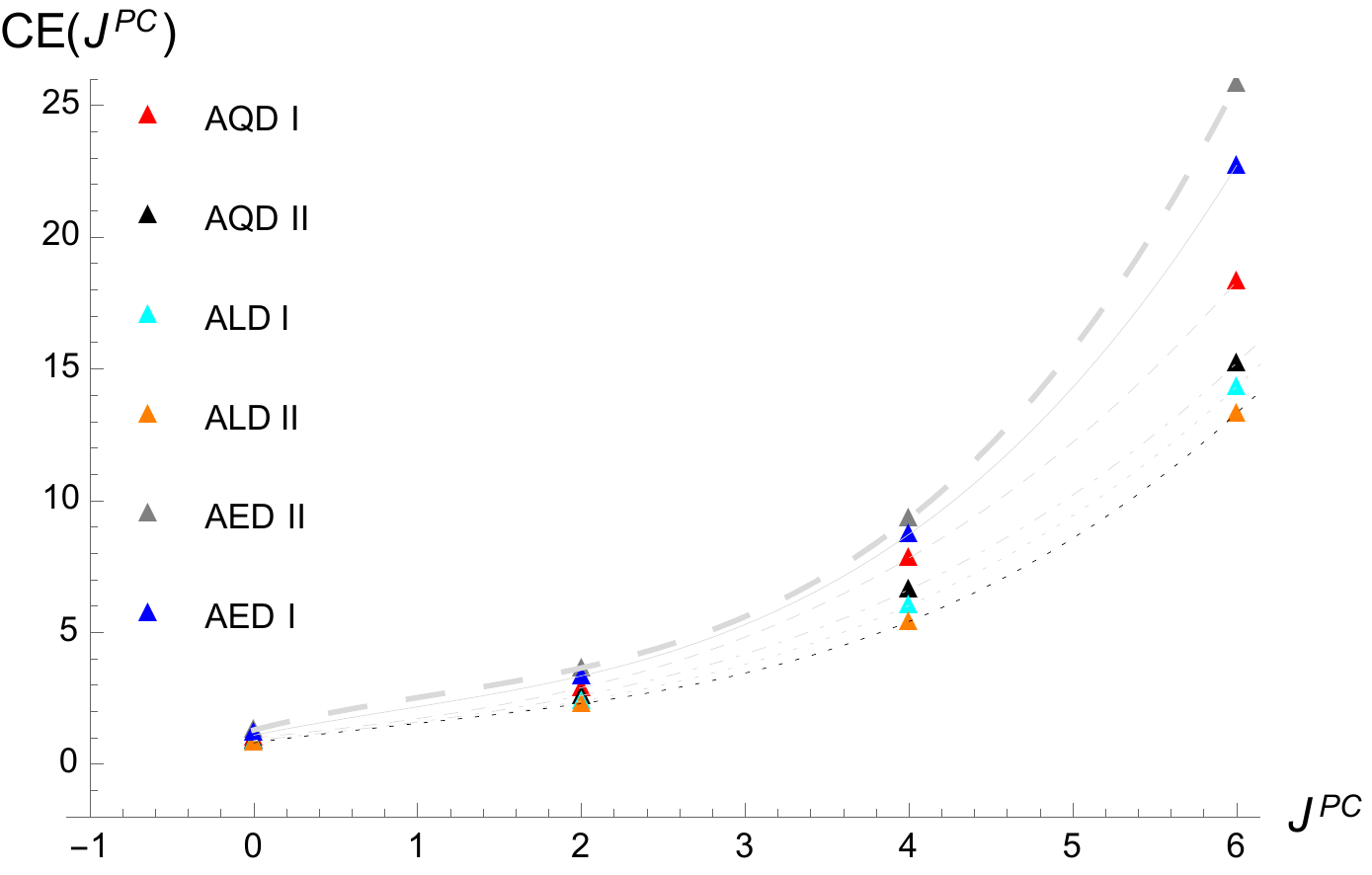}
\caption{CE of the \blt{even-spin glueball} family with respect to $J^{PC}$, taking into account exclusively lattice QCD data, up to $J^{PC}=6$. The anomalous quadratic dilaton I [II] model is shown by red [black] triangles, whereas the anomalous exponential dilaton I [II] model is illustrated by blue [gray] triangles; and the anomalous linear dilaton I [II] model is depicted by cyan [orange] triangles. }
\label{figcexjmenos}
\end{figure}

For the anomalous quadratic dilaton I model, the first type of CE Regge trajectory, when one considers exclusively the available lattice QCD data up to $J^{PC}=6$ from lattice QCD, reads 
\begin{eqnarray}
 {\rm CE}_{\scalebox{.55}{AQD I}}(J^{PC}) &=& 0.05333\, \left({J^{PC}}\right)^3 +0.05750\, \left({J^{PC}}\right)^2 \nonumber\\&&+0.62166\, {J^{PC}}+ 0.99999.\label{itp1menos}
  \end{eqnarray} It is the dashed light-gray curve in Fig. \ref{figcexjmenos}, that interpolates red triangles with accuracy within $0.11\%$. 
Besides, the expression of the CE Regge trajectory for the anomalous exponential dilaton I  model, representing the continuous curve, that interpolates blue triangles in Fig. \ref{figcexjmenos}, is given by
\begin{eqnarray}
 {\rm CE}_{\scalebox{.55}{AED I}}(J^{PC}) &=& 0.11479\, \left({J^{PC}}\right)^3 -0.29874\, \left({J^{PC}}\right)^2 \nonumber\\&&+1.26333\, {J^{PC}}+ 1.09999.\label{itp1expmenos}
  \end{eqnarray}
  \noindent 
  The formula of the CE Regge trajectory, for the anomalous linear dilaton I model, reads
\begin{eqnarray}
 {\rm CE}_{\scalebox{.55}{ALD I}}(J^{PC}) &=& 0.05352\, \left({J^{PC}}\right)^3 -0.05850\, \left({J^{PC}}\right)^2 \nonumber\\&&+0.66241\, {J^{PC}}+ 0.88099.\label{itp1linmenos}
  \end{eqnarray} It corresponds to the dotted curve, interpolating cyan triangles in Fig. \ref{figcexj}, with accuracy within $0.21\%$. 
  
  The expression of the CE Regge trajectory, for the anomalous quadratic dilaton II model, reads
\begin{eqnarray}
 {\rm CE}_{\scalebox{.55}{AQD II}}(J^{PC}) &=& 0.046875\, \left({J^{PC}}\right)^3 +0.008749\, \left({J^{PC}}\right)^2 \nonumber\\&&+0.645\, {J^{PC}}+ 0.900.\label{itp1logmenos}
  \end{eqnarray} It corresponds to the light-gray dot-dashed curve, interpolating black triangles in Fig. \ref{figcexj}. Cubic interpolation suffices to delimit accuracy within $\sim0.19\%$. 
    \noindent 
  Besides, the expression of the CE Regge trajectory for the anomalous exponential dilaton II  model, representing the large-dashed light-gray interpolating curve in Fig. \ref{figcexjmenos}, is given by
\begin{eqnarray}
 {\rm CE}_{\scalebox{.55}{AED II}}(J^{PC}) &=& 0.15625\, \left({J^{PC}}\right)^3 -0.5225\, \left({J^{PC}}\right)^2 \nonumber\\&&+1.5950\, {J^{PC}}+ 1.300.\label{itp1expmenosii}
  \end{eqnarray} It has accuracy of $0.13\%$.
  \noindent 
  Finally, the formula of the CE Regge trajectory, for the anomalous linear dilaton II model, reads
\begin{eqnarray}
 {\rm CE}_{\scalebox{.55}{ALD II}}(J^{PC}) &=& 0.0660208\, \left({J^{PC}}\right)^3 -0.1935\, \left({J^{PC}}\right)^2 \nonumber\\&&+ 0.86741\, {J^{PC}}+ 0.81100.\label{itp1linmenosii}
  \end{eqnarray} It corresponds to the black dotted curve, interpolating orange triangles, in Fig. \ref{figcexj}, with accuracy within $\sim0.8\%$.

\begin{figure}[h]
\centering
\includegraphics[width=8cm]{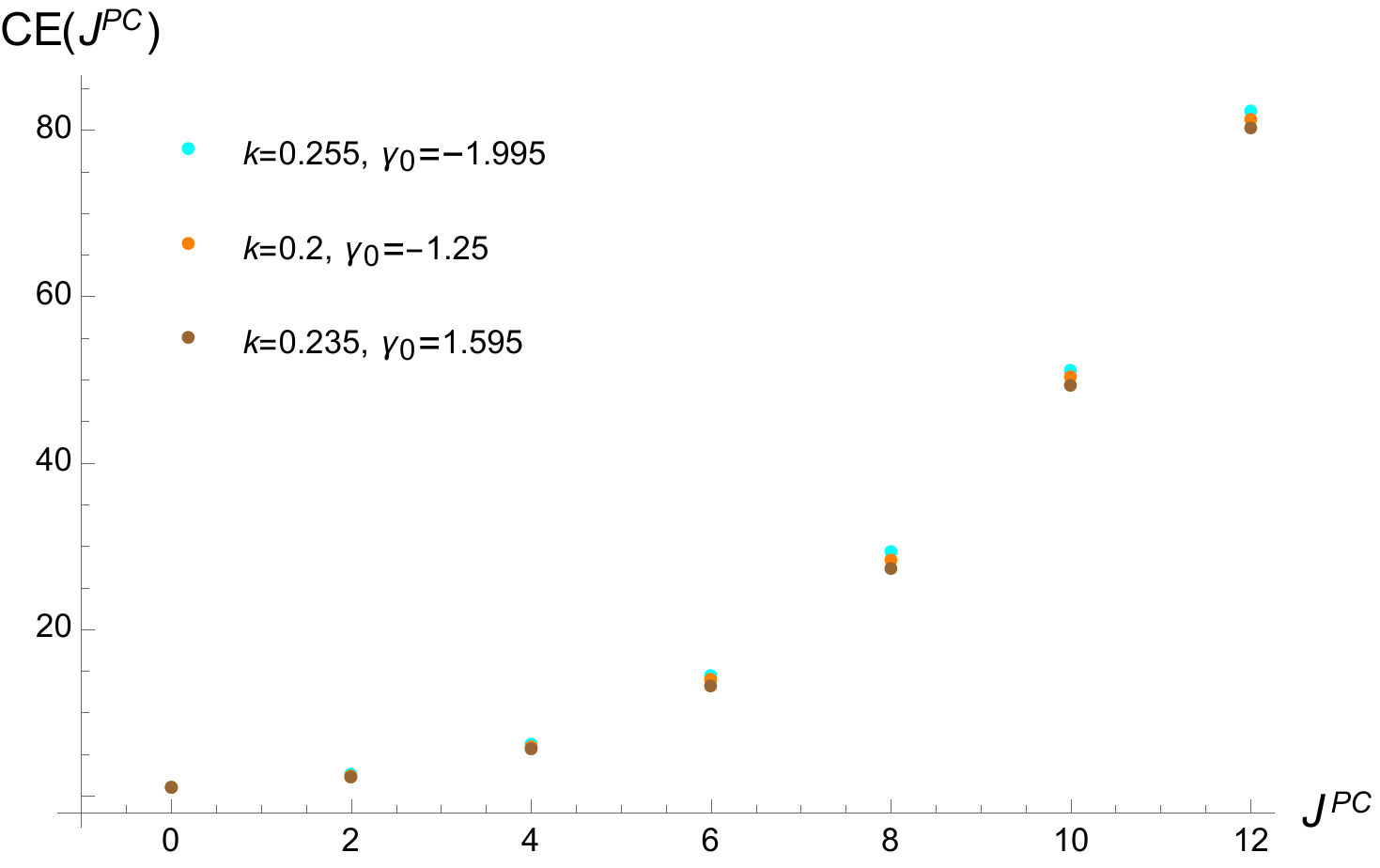}
\caption{CE of the \blt{even-spin glueball} family with respect to $J^{PC}$, for different values of $k$ and $\gamma_0$, in the anomalous linear model I. Cyan points represent $k=0.255$ and $\gamma_0=-1.995$; orange 
points depict the case $k=0.2$ and $\gamma_0=-1.25$; and brown points represent $k=0.235$ and $\gamma_0=1.595$.}
\label{figcexjlin}
\end{figure}

  It is worth to mention that, for the anomalous linear models, there are several possibilities that best fit 
 the QCD lattice \blt{even-spin glueballs} mass spectra, for low $J^{PC}$. Hence, some appropriate choices for 2-tuple $(k, \gamma_0)$ can be employed for matching QCD lattice predictions regarding the \blt{even-spin glueball} family. Some values best fit the $J^{PC}=0^{++}, 2^{++}$ \blt{even-spin glueballs}, whereas other values of $(k, \gamma_0)$ best fit higher $J^{PC}$ \blt{even-spin glueballs} resonances. For different values of $k$ and $\gamma_0$, the CE of the \blt{even-spin glueballs} with respect to $J^{PC}$ is plot in Fig. \ref{figcexjlin}.
\noindent Since the three appropriate choices of $(k, \gamma_0)$, studied and illustrated in Fig. \ref{figcexjlin}, generate the respective CEs, with a maximum variance of $1.1\%$, it indicates that they present approximately equal configurational stability. Hence, this is the reason we chose $k=0.255$ and $\gamma_0=-1.995$ in all the analysis regarding the anomalous linear dilaton I model.

When one attributes values $J^{PC}=8,10,\ldots,18$ in the CE Regge trajectories (\ref{itp1menos}, \ref{itp1expmenos}, \ref{itp1logmenos}), respectively for the six dilaton models, the CE of the \blt{even-spin glueball} family are compiled, in the second to the seventh columns of Table \ref{CEJmenos}.
\begin{table}[H]
\begin{center}
\begin{tabular}{||c||c||c||c||c||c||c||c||c||}
\hline\hline
   $J^{PC}$ &~CE$_{\scalebox{.53}{AQD I}}$& CE$_{\scalebox{.53}{AED I}}$& CE$_{\scalebox{.53}{ALD I}}$&~CE$_{\scalebox{.53}{AQD II}}$&~CE$_{\scalebox{.53}{AED II}}$&~CE$_{\scalebox{.53}{ALD II}}$ \\\hline\hline
   \, 0 \,&$1.00$&1.10&0.88&0.92&1.30&0.81 \\\hline
   \, 2 \,&$2.99$&3.35&2.30&2.62&3.66&2.43\\\hline
   \, 4 \,&$7.82$&8.72&5.42&6.62&9.32&6.02\\\hline
   \, 6\, &$18.32$&22.73&13.32&15.21&25.81&14.31 \\\hline
   \, 8\, &$36.96$&50.86&28.97&30.62&60.62&27.41\\\hline
   \, 10 \,&$66.34$&98.65&50.98&55.12&121.25&48.99 \\\hline
    \, 12 \,&$108.93$&171.63&82.14&90.94&215.20&79.94 \\\hline
     \, 14 \,&$167.32$&275.22&124.05&140.27&349.97&121.99 \\\hline
      \, 16 \,&$244.13$&415.02&178.26&205.46&533.06&176.83\\\hline
      \, 18 \,&$341.87$&596.52&246.38&288.72&771.97&246.16
      \\\hline\hline
\end{tabular}\caption{The CE of \blt{even-spin glueballs} as a function of $J^{PC}$, taking into account exclusively lattice QCD data up to $J^{PC}=6$. The second and fifth columns display the CE of the anomalous quadratic dilaton I and II (AQD I and II) models; the third and sixth columns show the \blt{even-spin glueballs} CE computed from the anomalous exponential dilaton I and (AED I and II) models; the fourth and seventh columns show the CE, respectively computed from the anomalous linear dilaton I and II (ALD I and II) models. }
\label{CEJmenos}
\end{center}
\end{table}

Similarly for the hybrid model shown in Fig. \ref{figcexj}, also Fig. \ref{figcexjmenos} illustrates that the anomalous exponential dilaton I  model regards \blt{even-spin glueballs} resonances that are configurationally more unstable states when compared with other dilaton models. Again, the anomalous linear dilaton I model does designate \blt{even-spin glueballs} resonances that are more stable, from the configurational point of view. 

\begin{figure}[h]
\centering
\includegraphics[width=8cm]{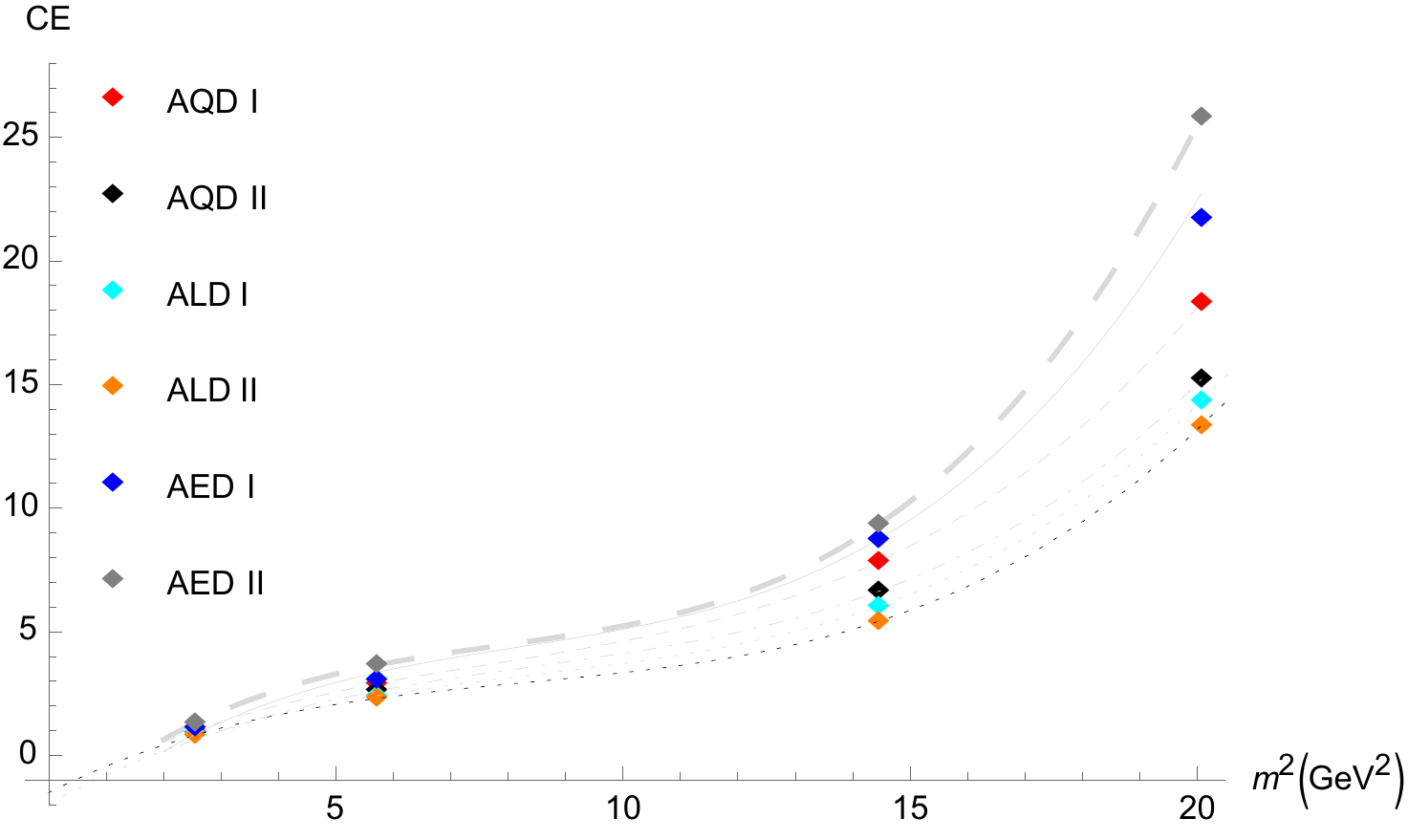}
\caption{CE of the \blt{even-spin glueball} family with respect to their mass spectra, taking into account exclusively lattice QCD data, up to $J^{PC}=6$. The anomalous quadratic dilaton I [II] model is shown by the red [black] diamonds, whereas the anomalous exponential dilaton I [II] model is illustrated by the blue [gray] diamonds; and the anomalous linear dilaton I [II] model is depicted by cyan [orange] diamonds. }
\label{figm2xjmenos}
\end{figure}

Now, with the \blt{even-spin glueball} family mass spectra, for $J^{PC}=0,2,4,6$, obtained from lattice QCD, higher spin \blt{even-spin glueballs} resonances can have their masses determined, when the second type of CE Regge trajectories is considered.
They interpolate the plot of the \blt{even-spin glueball} family, for $J^{PC}=0,2,4,6$, with respect to their mass spectra, for the six dilaton models, and displayed in Fig. \ref{figm2xjmenos}.
 Higher $J^{PC}$ \blt{even-spin glueballs} resonances will have their mass spectrum, therefore, extrapolated from Eqs. (\ref{itp1menos}, \ref{itq1menos}), for the anomalous quadratic dilaton I model; from Eqs. (\ref{itp1expmenos}, \ref{itq1expmenos}), for the anomalous exponential dilaton I  model; from Eqs. (\ref{itp1logmenos}, \ref{itq1logmenos}), for the anomalous quadratic dilaton model; and from Eqs. (\ref{itp1linmenos}, \ref{itq1linmenos}) and (\ref{itp1linmenosii}, \ref{itq1linmenosii}), respectively for the anomalous linear dilaton I and II  models.
 
Therefore the CE of \blt{even-spin glueballs} with respect to the (squared) mass spectra, for the anomalous quadratic dilaton I model, is shown by the dashed curve in Fig. \ref{figm2xjmenos}, that interpolates the red diamonds, reading 
\begin{eqnarray}\label{itq1menos}
\!\!\!\!\!\!\!\! {\rm CE}_{\scalebox{.55}{AQD I}}(m) \!&=\!&\! 0.00534334 m^6\!-\!0.1243014m^4\!\nonumber\\&&+1.339408\,m^2-1.690979,  \end{eqnarray} within $0.78\%$ standard deviation. 
For obtaining the mass spectra of the higher spin resonances in the \blt{even-spin glueball} family, let us first analyze the CE Regge trajectory \eqref{itq1menos}, also derived from the anomalous quadratic dilaton I model. In fact, choosing for $J^{PC}=8$, Eq. (\ref{itp1menos}) yields 
${\rm CE} = 36.961$. When this value is replaced in the CE Regge trajectory (\ref{itq1menos}), it yields the mass 
\beq
m_{\mathcal{P}_{8}}=4.9875\; {\rm GeV}
\eeq It yields the range $5.00\; {\rm GeV}\lesssim m_{\mathcal{P}_{8}}\lesssim 4.96\;{\rm GeV}$. Analogously, when one substitutes $J^{PC}=10$ into Eq. (\ref{itp1menos}), it yields the CE equal to $66.30$. Putting back in Eq. (\ref{itq1menos}) produces \beq
m_{\mathcal{P}_{10}}=5.4253\; {\rm GeV,}\eeq
 being the mass range $5.44 \,{\rm GeV}\lesssim m_{\mathcal{P}_{10}}\lesssim 5.48\, {\rm GeV}$. For the $\mathcal{P}_{12}$ \blt{even-spin glueball}, the CE Regge trajectory (\ref{itp1menos}) yields ${\rm CE}_{\mathcal{P}_{12}} = 108.98$, when $J^{PC}=12$ is taken into account. When reinstated in Eq. (\ref{itq1menos}), the mass 
 \beq
 m_{\mathcal{P}_{12}}=5.8214\; {\rm GeV}
 \eeq is then derived. Similarly, replacing $J^{PC}=14$ in Eq. (\ref{itp1menos}) implies ${\rm CE}_{\mathcal{P}_{14}} = 167.32$. Consequently, when one solves Eq.(\ref{itq1menos}) for this value, it implies that 
 \beq\label{m18log33}
 m_{\mathcal{P}_{14}}=6.1879\; {\rm GeV}. 
\eeq
In the same way, the CE Regge trajectory (\ref{itp1menos}) yields ${\rm CE}_{\mathcal{P}_{16}} = 244.12$, when $J^{PC}=16$ is taken into account. When reinstated in Eq. (\ref{itq1menos}), the mass 
 \beq
 m_{\mathcal{P}_{16}}=6.5316\; {\rm GeV}
 \eeq is derived. Moreover, the value $J^{PC}=18$ is associated to 
 ${\rm CE}_{\mathcal{P}_{18}} = 341.86$ that can be replaced in Eq. (\ref{itq1menos}), producing  
 \beq
 m_{\mathcal{P}_{18}}=6.8568\; {\rm GeV}.
 \eeq 
These results are compiled in the second column in Table \ref{CEM2menos}.

The CE Regge trajectory, that interpolates the CE of the \blt{even-spin glueball} family, for $J^{PC}=0,2,4,6$, to the \blt{even-spin glueball} (squared) mass spectra, is shown by the continuous curve in Fig. \ref{figm2xjmenos}, for the anomalous exponential dilaton I, 
\begin{eqnarray}\label{itq1expmenos}
\!\!\!\!\!\!\!\! {\rm CE}_{\scalebox{.55}{AED I}}(m) \!&=\!&\! 0.00829209\; m^6\!-\!0.203204\;m^4\!\nonumber\\&&+2.026744\,m^2-3.142225,  \end{eqnarray} within $0.4\%$ standard deviation. 
For the anomalous exponential dilaton I  model, in fact, choosing for $J^{PC}=8$, Eq. (\ref{itp1expmenos}) yields 
${\rm CE} = 50.863$. When this value is replaced in the CE Regge trajectory (\ref{itq1expmenos}), it yields the mass 
\beq
m_{\mathcal{P}_{8}}=5.0103\; {\rm GeV,}
\eeq for the $\mathcal{P}_{8}$ \blt{even-spin glueball}. It yields the range $4.99\; {\rm GeV}\lesssim m_{\mathcal{P}_{8}}\lesssim 5.03\;{\rm GeV}$. \cclt{Doing the same calculations analogously, for $J^{PC}=10,\ldots,18$, the derived mass spectra for this dilaton model is displayed in the third column of Table \ref{CEM2menos}.}

Respectively, for the anomalous linear dilaton I  model, 
the CE values of the \blt{even-spin glueball} family, for $J^{PC}=8,10,\ldots,18$ are displayed in the fourth column of Table \ref{CEM2menos}. The CE Regge trajectory, interpolating the CE to the \blt{even-spin glueballs} (squared) mass spectra, is displayed the dotted curve in Fig. \ref{figm2xjmenos}, being given by 
\begin{eqnarray}\label{itq1linmenos}
\!\!\!\!\!\!\!\! {\rm CE}_{\scalebox{.55}{ALD I}}(m) \!&=\!&\! 0.0048877\; m^6\!-\!0.12293\;m^4\!\nonumber\\&&+1.31044\,m^2-1.98507,  \end{eqnarray} within $0.1\%$ standard deviation. Choosing for $J^{PC}=8$, Eq. (\ref{itp1linmenos}) yields 
${\rm CE} = 28.9767$. Now, replacing this value in (\ref{itq1linmenos}) yields the mass 
\beq
m_{\mathcal{P}_{8}}=4.96854\; {\rm GeV,}
\eeq for the $\mathcal{P}_{8}$ \blt{even-spin glueball}. It yields the range $4.94\; {\rm GeV}\lesssim m_{\mathcal{P}_{8}}\lesssim 4.98\;{\rm GeV}$. \cclt{One can accomplish similar computations, for $J^{PC}=10,\ldots,18$, obtaining the derived mass spectra for this dilaton model in the fourth column of Table \ref{CEM2menos}.}

Besides, the CE Regge trajectory, interpolating the CE to the \blt{even-spin glueballs} (squared) mass spectra, is displayed the dot-dashed curve in Fig. \ref{figm2xjmenos}, for the anomalous quadratic II dilaton, 
\begin{eqnarray}\label{itq1logmenos}
\!\!\!\!\!\!\!\! {\rm CE}_{\scalebox{.55}{AQD II}}(m) \!&=\!&\! 0.0050079\; m^6\!-\!0.12725\;m^4\!\nonumber\\&&+1.40438\,m^2-2.20316,  \end{eqnarray} within $0.2\%$ standard deviation. 
 In fact, choosing for $J^{PC}=8$, Eq. (\ref{itp1logmenos}) yields 
${\rm CE} = 30.621$. When this value is replaced in the CE Regge trajectory (\ref{itq1logmenos}), it yields the mass 
\beq
m_{\mathcal{P}_{8}}=4.9771\; {\rm GeV,}
\eeq for the $\mathcal{P}_{8}$ \blt{even-spin glueball}. It yields the range $4.94\; {\rm GeV}\lesssim m_{\mathcal{P}_{8}}\lesssim 4.99\;{\rm GeV}$. \cclt{The results for $J^{PC}=10,\ldots,18$ are listed in the fifth column in Table \ref{CEM2menos}.}

The CE Regge trajectory, that interpolates the CE of the \blt{even-spin glueball} family, for $J^{PC}=0,2,4,6$, to the \blt{even-spin glueballs} (squared) mass spectra, is shown by the continuous curve in Fig. \ref{figm2xjmenos}, for the anomalous exponential dilaton II, 
\begin{eqnarray}\label{itq1expmenosii}
\!\!\!\!\!\!\!\! {\rm CE}_{\scalebox{.55}{AED II}}(m) \!&=\!&\! 0.00949858\; m^6\!-\!0.223326\;m^4\!\nonumber\\&&+2.07616\,m^2-2.69281,  \end{eqnarray} within $0.4\%$ standard deviation. 
For the anomalous exponential dilaton II  model, in fact, choosing for $J^{PC}=8$, Eq. (\ref{itp1expmenosii}) yields 
${\rm CE} = 60.621$. When this value is replaced in the CE Regge trajectory (\ref{itq1expmenosii}), it yields the mass 
\beq
m_{\mathcal{P}_{8}}=5.0289\; {\rm GeV,}
\eeq for the $\mathcal{P}_{8}$ \blt{even-spin glueball}. It yields the range $4.99\; {\rm GeV}\lesssim m_{\mathcal{P}_{8}}\lesssim 5.03\;{\rm GeV}$. \cclt{The derived mass spectra of the \blt{even-spin glueball} family, for the further states $J^{PC}=10,\ldots,18$, is displayed in the sixth column of Table \ref{CEM2menos}}.

Respectively, for the anomalous linear dilaton II  model, 
the CE values of the \blt{even-spin glueball} family, for $J^{PC}=8,10,\ldots,18$ are displayed in the fourth column of Table \ref{CEJmenos}. The CE Regge trajectory, relating the \blt{even-spin glueballs} mass spectra to the CE, for this model, reads
\begin{eqnarray}\label{itq1linmenosii}
\!\!\!\!\!\!\!\! {\rm CE}_{\scalebox{.55}{ALD II}}(m) \!&=\!&\! 0.00470485\; m^6\!-\!0.116337\;m^4\!\nonumber\\&&+1.17817\,m^2-1.51081,  \end{eqnarray} within $0.4\%$ standard deviation. 
\cclt{Consequently, employing the same procedure using the CE Regge trajectories (\ref{itp1linmenosii}, \ref{itq1linmenosii}), the derived mass spectra of the \blt{even-spin glueball} family, for $J^{PC}=8,10,\ldots,18$, is displayed in the seventh column of Table \ref{CEM2menos}.}
\begin{table}[H]
\begin{center}\medbreak
\begin{tabular}{||c||c||c||c||c||c||c||c||c||}
\hline\hline
   $J^{PC}$ & $m_{\scalebox{.55}{AQD I}}$& $m_{\scalebox{.55}{AED I}}$&$m_{\scalebox{.55}{ALD I}}$&$m_{\scalebox{.55}{AQD II}}$& $m_{\scalebox{.55}{AED II}}$&$m_{\scalebox{.55}{ALD II}}$ \\\hline\hline
 \, 8 \,&4.99&5.01&4.97&4.98&5.03&4.97 \\\hline
 \, 10 \,&5.42&5.48&5.37&5.40&5.52&5.38 \\\hline
    \, 12 \,&5.82&5.90&5.74&5.79&5.96&5.75 \\\hline
     \, 14 \,&6.19&6.30&6.07&6.15&6.37&6.09 \\\hline
      \, 16 \,&6.53&6.67&6.39&6.49&6.76&6.41\\\hline
      \, 18 \,&6.86&7.02&6.68&6.81&7.13&6.71 \\\hline\hline
\end{tabular}\caption{Mass spectra (GeV) of \blt{even-spin glueballs} as a function of $J^{PC}$, taking into account exclusively lattice QCD data up to $J^{PC}=6$. The second and fifth columns display the CE of the anomalous quadratic dilaton I and II (AQD I and II) models; the third and sixth columns show the \blt{even-spin glueballs} CE computed from the anomalous exponential dilaton I and (AED I and II) models; the fourth and seventh columns show the CE, respectively computed from the anomalous linear dilaton I and II (ALD I and II) models. }
\label{CEM2menos}
\end{center}
\end{table}

One can compare the \blt{even-spin glueballs} mass spectra in Tables \ref{CEM2} and \ref{CEM2menos}, computed from the CE Regge trajectories, with the ones
purely computed from AdS/QCD in Tables \ref{CEJqcdlin} and \ref{CEJqcdlog} .

\begin{table}[H]
\begin{center}------------------------------------------------------------------------------\\ \blt{Even-spin glueballs} mass spectra: anomalous linear dilaton I model \medbreak
\begin{tabular}{||cc||c||c||c||c||c||}
\hline\hline
 &  $J^{PC}$ &~$k=0.17$~&~$k=0.16$~&~$k=0.13$~&~$k=0.255$~\\
 &&~$\gamma_0=-1.0$~&$~\gamma_0=-1.5$~&~$\gamma_0=-1.0$~&~$\gamma_0=-1.995$~ \\\hline\hline
   \,&\, 8\, &$5.073$&4.813&4.494&5.837\\\hline
   \,&\, 10 \,&$6.177$&5.882&5.464&7.176 \\\hline
    \,&\, 12 \,&$7.281$&6.952&6.435&8.519 \\\hline
     \,&\, 14 \,&$8.385$&8.023&7.405&9.863 \\\hline
      \,&\, 16 \,&$9.489$&9.095&8.376&11.209\\\hline
      \,&\, 18 \,&$10.594$&10.167&9.347&12.555 \\\hline\hline
\end{tabular}
\caption{\blt{Even-spin glueballs} mass spectra (GeV) as a function of $J^{PC}$, for the anomalous linear dilaton I model, from AdS/QCD, for several values of $k$ and $\gamma_0$. }
\label{CEJqcdlin}
\end{center}
\end{table}
The fourth column in Table \ref{CEM2menos}, for the values $k=0.255$ and $\gamma_0=-1.995$, represents the \blt{even-spin glueballs} mass spectra obtained purely from AdS/QCD.
These values were used to derive the CE Regge trajectories (\ref{itp1lin}, \ref{itq1lin}) --- for the hybrid model considering exclusively lattice QCD data, up to $J^{PC}=6$, and including for $J^{PC}=8, 10$ the \blt{even-spin glueballs} mass spectra (\ref{massa1}) and extrapolating for $J^{PC}=12,\ldots,18$ the mass spectra --- and (\ref{itp1linmenos}, \ref{itq1linmenos}) --- regarding exclusively lattice QCD data, up to $J^{PC}=6$, and extrapolating for $J^{PC}=8,\ldots,18$ the mass spectra. Thence, the values $k=0.255$ and $\gamma_0=-1.995$ have been employed to compute the CE with the respect to both $J^{PC}$ and the \blt{even-spin glueballs} mass spectra, represented by cyan [orange] points for the anomalous linear dilaton I [II], respectively in Figs. \ref{figcexj} and \ref{figcexm2}. Let us compare the \blt{even-spin glueballs} mass spectra, generated from the anomalous linear dilaton I model in the hybrid model in Table \ref{CEM2}, to the \blt{even-spin glueballs} mass spectra obtained purely from AdS/QCD in Table \ref{CEJqcdlin}. Let us denote by $\Updelta_k$ the difference between the respectively derived masses when $J^{PC}=k$. Therefore, when $J^{PC}=12$, $\Updelta_{12}=16.5\%$, whereas the value $J^{PC}=14$ yields $\Updelta_{12}=14.6\%$. Also, $\Updelta_{16}=24.3\%$ for $J^{PC}=16$ and $\Updelta_{18}=34.7\%$. The increment of the mass difference, as $k$ increases, is understood as the uncertainty in the models for higher \blt{even-spin glueballs} resonances. A similar reasoning can be obtained for all other dilaton models. 

\begin{table}[H]
\begin{center}-----------------------------------------------------------------------\\ \blt{Even-spin glueballs} mass spectra: anomalous linear II model \medbreak
\begin{tabular}{||cc||c||c||c||c||c||}
\hline\hline
 &  $J^{PC}$ &~$k=0.2$~&~$k=0.2$~&~$k=0.12$~\\
 &&~$\gamma_0=-1.5$~&$~\gamma_0=1.595$~&~$\gamma_0=1.233$~ \\\hline\hline
   \,&\, 8\, &$5.626$&5.841&4.605\\\hline
   \,&\, 10 \,&$6.838$&7.032&5.538 \\\hline
    \,&\, 12 \,&$8.046$&8.223&6.471 \\\hline
     \,&\, 14 \,&$9.252$&9.415&7.404 \\\hline
      \,&\, 16 \,&$10.457$&10.607&8.337\\\hline
      \,&\, 18 \,&$11.660$&11.800&9.271 \\\hline\hline
\end{tabular}
\caption{\blt{Even-spin glueballs} mass spectra (GeV) as a function of $J^{PC}$, for the anomalous linear dilaton II model, from AdS/QCD. }
\label{CEJqcdlog}
\end{center}
\end{table}

\section{Concluding remarks}\label{sec3}

The CE was computed for \blt{even-spin glueballs}, for six dilaton models, namely, the anomalous quadratic dilaton I and II, the anomalous linear dilaton I and II, and the anomalous exponential dilaton I and II models. Two types of configurational-entropic Regge trajectories were derived, for the \blt{even-spin glueballs} resonances. 
The first CE Regge trajectory relates the CE of the \blt{even-spin glueballs} to their $J^{PC}$ spin, represented by Eqs. (\ref{itp1} -- \ref{itp1linii}), respectively for the anomalous quadratic I, the anomalous exponential dilaton I, the anomalous linear I, the anomalous quadratic II, the anomalous exponential II, and the anomalous linear II dilaton models. These CE Regge trajectories are interpolation curves, taking into account exclusively lattice QCD data, up to $J^{PC}=6$, and including for $J^{PC}=8, 10$ the \blt{even-spin glueballs} mass spectra (\ref{massa1}), shown in Fig. \ref{figcexj}. The second type of CE Regge trajectories 
relates the CE of \blt{even-spin glueballs} with their mass spectra, for the six dilaton models, represented by Eqs. (\ref{itq1}, \ref{itq1exp}, \ref{itq1lin}, \ref{itq1log}, \ref{itq1expii}, \ref{itq1linii}), respectively shown in the plots of Fig. \ref{figcexm2}. Consequently, the \blt{even-spin glueball} family mass spectra were extrapolated  
from the CE Regge trajectories. A range for the mass spectra  
of \blt{even-spin glueballs} with higher $J^{PC}$ was then estimated with good accuracy. 
Higher \blt{even-spin glueballs} resonances, beyond $J^{PC}=18$, are not explored, since 
they are configurationally very unstable, with huge values of CE.
Even states with $J^{PC}>12$ are unlikely to be experimentally detected, at least with proposed experiments that have been run in LHC. 

For both the protocols adopted, a) in Subsect. \ref{chan1} computing the CE of \blt{even-spin glueballs} as a function of $J^{PC}$, taking into account lattice QCD data up to $J^{PC}=6$, and including for $J^{PC}=8, 10$ the \blt{even-spin glueballs} mass spectra (\ref{massa1}); and b) in Subsect. \ref{chan2}, taking into account exclusively lattice QCD data, up to $J^{PC}=6$ --- both Figs. \ref{figcexj} and \ref{figcexjmenos} exhibit the \blt{even-spin glueballs} resonances in the anomalous exponential dilaton I as more unstable, from the configurational point of view, when compared with the other dilaton models approached. Besides, the linear dilaton model II encompasses more stable \blt{even-spin glueballs} resonances. More configurationally stable resonances represent quantum states that are more dominant and prevalent, from the phenomenological standpoint \cite{Bernardini:2016hvx,Bernardini:2018uuy,Bernardini:2016qit}. Hence, this method points to \blt{even-spin glueballs} resonances that may be more detectable and observable in experiments. 

From the experimental point of view, the $f_2(1950)$ resonance, with $J^{PC}=2^{++}$ and mass $1.944\pm 0.012$ GeV \cite{pdg1}, has been recently proposed to 
be a promising candidate for the \blt{even-spin glueball} ground state \cite{Godizov:2016vuw}.
Since there is no experimental data available in PDG for high $J^{PC}$ \blt{even-spin glueballs} resonances, then at least the obtained mass spectra should be compared to lattice QCD data. However, available calculations encompass just low $J^{PC}$ \blt{even-spin glueballs} resonances. 
Hence, the analysis and tables heretofore obtained in this work, containing the \blt{even-spin glueballs} mass spectra for the six dilaton models, allows us to create a quite useful phenomenological database. The CE Regge trajectories provide the derivation of the mass spectra of higher $J^{PC}$ \blt{even-spin glueballs} resonances, acting as a benchmarking of AdS/QCD models.

\aalt{One of
the advantages of employing CE-based techniques is the 
the straightforward computational tools, when one compares 
to other approaches of AdS/QCD, comprising lattice QCD. Another goal of using the CE for even-spin glueballs is the interpolation of a mixture of AdS/QCD and lattice QCD, as implemented in Section \ref{chan1}, to derive the mass spectra of the next generation of even-spin glueball resonances.  The lower the information
entropy of a system, the lower the chaoticity it presents. Besides, as the CE is equivalent to the information chaoticity carried by a message, one can shed light on the even-spin glueball families as results of experimental particle production processes. In high-energy collisions involving gluons and quarks, chaos in QCD gauge dynamics may set in, and only particles in final states can be measured. The chaotic profile
of collisions that produce even-spin glueball excitations can
be then quantified by the loss of information at the end
of the collision processes. This quantitative aspect was
here shown to be encompassed by the CE, providing
phenomenological determination by the derived mass spectra
of even-spin glueball states with higher excitation numbers. This additional interpretation complies to other results in the literature, regarding chaos in QCD and holography (see, e. g., \cite{PandoZayas:2010xpn,Hashimoto:2016wme,Pascalutsa:2002kv, Muller:1992iw,Basu:2013uva,Maldacena:2015waa}).
Tables \ref{CES} -- \ref{CES5}, for all dilaton models under scrutiny,  compared the obtained mass spectra, in the CE approach, to the ones derived by lattice QCD and also the DP Regge model,  for $J \leq 6$ and $J\leq 8$, respectively. One more relevant point of the CE approach is that the DP Regge model and lattice QCD
do not derive  the mass spectra for $J>8$. Above it, there is no data in the literature, up to our knowledge. 
One of our aims is  to provide and make available a useful phenomenological database, presenting a data bank for high-energy reactions and decays, having even-spin glueballs as a byproduct in high-energy experiments.  Since the odderon has been recently detected \cite{TOTEM:2020zzr,Csorgo:2019ewn} at TOTEM LHC and DØ experiment at Tevatron, we expect that data about even-spin glueballs, including the Pomeron, may be soon also evinced experimentally.   The comparison between 
 AdS/QCD-obtained mass spectra and the CE-based procedure that here hybridizes AdS/QCD should be also analyzed with  available experimental data, of the higher excitations that are still beyond the current detection threshold.  }

Furthermore, it would be interesting to extend the analysis made in this paper
considering different profiles for the dilaton field, and also considering glueball operators of twist-2, since some works propose and consider the pomeron to be associated with a twist-2 object \cite{Ewerz:2013kda,Iatrakis:2016rvj,Szanyi:2019kkn}. Here, we considered the twist-4 approach to include the scalar glueball sector. However, there are some shreds of evidence, within the holographic approach, which show that the mass of the scalar and tensor glueballs are degenerate. Here, we provided one more piece of evidence for this in the case of the exponential dilaton profile. Finally, another possibility is to extend the analysis we have made here to the odd-spin glueballs, which are associated with the odderon.

\medbreak
\paragraph*{Acknowledgments:} We would like to thank Song He for useful correspondence. DMR is supported by the National Council for Scientific and Technological Development -- CNPq (Brazil) under Grant No. 152447/2019-9 and to FAPESP (Grant No.	2021/01565-8). RdR~is grateful to The S\~ao Paulo Research Foundation - FAPESP (Grants No. 2017/18897-8, No. 2021/01089-1, and No. 2022/01734-7) and the National Council for Scientific and Technological Development -- CNPq (Grants No. 303390/2019-0  and No. 402535/2021-9), for partial financial support.   
\medbreak
\paragraph*{Data Availability Statements:} the datasets generated during and/or analysed during the current study are available from the corresponding author on reasonable request.

\end{document}